\documentclass[12pt]{iopart}
\usepackage{iopams}  
\usepackage[svgnames]{xcolor}
\usepackage[utf8]{inputenc}
\usepackage{dsfont}

\expandafter\let\csname equation*\endcsname\relax
\expandafter\let\csname endequation*\endcsname\relax

\usepackage{amsmath}
\usepackage{graphicx}
\usepackage{subfigure}
\usepackage{multirow}
\usepackage[linktocpage=true]{hyperref}
\hypersetup{
  colorlinks=true,
  citecolor=red,
  urlcolor=Blue}

\usepackage{ulem} 
\usepackage{units} 
\usepackage{version} 
\usepackage{color}
\usepackage{colortbl}
\usepackage{xcolor} 
\usepackage{comment}
\usepackage[utf8]{inputenc}
\usepackage[T1]{fontenc}
\usepackage{lmodern}
\usepackage{tabularx}
\usepackage{eurosym}

\usepackage[
  colorinlistoftodos, 
  textwidth=\marginparwidth, 
  textsize=scriptsize, 
  ]{todonotes}

\begin{document}

\title[Exploring gravity with the MIGA large scale atom interferometer]{Exploring gravity with the MIGA large scale atom interferometer}

\author{B. Canuel$^{1,2,*}$,
A. Bertoldi$^{1,2,*}$,
L. Amand$^{1,3}$,
E. Pozzo di Borgo$^{4}$,
T. Chantrait$^{1,3}$,
C. Danquigny$^{4}$,
M. Dovale Alvarez$^{5}$,
B. Fang$^{1,3}$,
A. Freise$^{5}$,
R. Geiger$^{1,3}$,
J. Gillot$^{1,2}$,
S. Henry$^{6}$,
J. Hinderer$^{1,7}$,
D. Holleville$^{1,3}$,
J. Junca$^{1,2}$
G. Lef{\`e}vre$^{1,2}$,
M. Merzougui$^{1,8}$,
N. Mielec$^{1,3}$,
T. Monfret$^{9}$,
S. Pelisson$^{1,2}$,
M. Prevedelli$^{10}$,
S. Reynaud$^{1,11}$,
I. Riou$^{1,2}$,
Y. Rogister$^{1,7}$,
S. Rosat$^{1,7}$,
E. Cormier$^{1,12}$,
A. Landragin$^{1,3}$,
W. Chaibi$^{1,8}$,
S. Gaffet$^{1,9,13}$,
and P. Bouyer$^{1,2}$}
\address{$^1$ MIGA Consortium}
\address{$^2$ LP2N, Laboratoire Photonique, Numérique et Nanosciences, Université Bordeaux--IOGS--CNRS:UMR 5298, rue F. Mitterrand, F--33400 Talence, France.}
\address{$^3$ LNE--SYRTE, Observatoire de Paris, Université PSL, CNRS, Sorbonne Université, 61, avenue de l’Observatoire, F--75014 PARIS, France}
\address{$^4$ EMMAH, UMR 1114 INRA/UAPV, Domaine Saint Paul, Site Agroparc, 228 route de l'Aérodrome CS 40509 84914 Avignon Cedex 9 France}
\address{$^{5}$ School  of  Physics  and  Astronomy  and  Institute  of  Gravitational  Wave  Astronomy,
University  of  Birmingham,  Edgbaston,  Birmingham  B15  2TT,  UK}
\address{$^{6}$ Oxford University, Department of Physics, Denys Wilkinson Building, Keble Road, Oxford, OX1 3RH UK}
\address{$^{7}$ Institut de Physique du Globe de Strasbourg, UMR 7516, Université de Strasbourg/EOST, CNRS,5 rue Descartes, 67084 Strasbourg, France}
\address{$^8$ ARTEMIS,  Université de Nice Sophia Antipolis, CNRS and Observatoire de la Côte d’Azur, F--06304 Nice, France}
\address{$^{9}$ GEOAZUR, UNS, CNRS, IRD, OCA , 250 rue Albert Einstein, 06560 Valbonne, France}
\address{$^{10}$ Dipartimento di Fisica e Astronomia, Universit{\`a} di Bologna, Via Berti-Pichat 6/2, I-40126 Bologna, Italy}
\address{$^{11}$ Laboratoire Kastler Brossel, CNRS, ENS-PSL Research University,  Collège de France, UPMC-Sorbonne Universités, Campus Jussieu, F-75252 Paris, France.}
\address{$^{12}$ CELIA, Centre Lasers Intenses et Applications, Université Bordeaux, CNRS, CEA, UMR 5107, F--33405 Talence, France}
\address{$^{13}$ LSBB, Laboratoire Souterrain à Bas Bruit, UNS, UAPV, CNRS:UMS 3538, AMU, La Grande Combe, F--84400 Rustrel, France.}

\address{$^{*}$ These authors contributed equally to this work}
\ead{benjamin.canuel@institutoptique.fr}
\vspace{5pt}

\begin{abstract}
We present an underground long baseline atom interferometer to study gravity at large scale. The hybrid atom-laser antenna will use several atom interferometers simultaneously interrogated by the resonant mode of an optical cavity. The instrument will be a demonstrator for gravitational wave detection in a frequency band (100 mHz -- 1 Hz) not explored by classical ground and space-based observatories, and interesting for potential astrophysical sources. In the initial instrument configuration, standard atom interferometry techniques will be adopted, which will bring to a peak strain sensitivity of 2$\cdot 10^{-13}/\sqrt{\mathrm{Hz}}$ at 2 Hz. The experiment will be realized at the underground facility of the Laboratoire Souterrain à Bas Bruit (LSBB) in Rustrel--France, an exceptional site located away from major anthropogenic disturbances and showing very low background noise. In the following, we present the measurement principle of an in-cavity atom interferometer, derive signal extraction for Gravitational Wave measurement from the antenna and determine the expected strain sensitivity. We then detail the functioning of the different systems of the antenna and describe the properties of the installation site.
\end{abstract}
\color{black}

%
%
%
%
%
\renewcommand{\footnoterule}{%
  \kern -3pt
  \hrule width \textwidth height 0.4pt
  \kern 2pt
}


\section{Introduction}

After its demonstration in 1991\cite{Carnal91, Keith91,Riehle91,Kasevich91}, atom interferometry (AI) rapidly revealed its huge potential for the precise measurement of inertial forces \cite{Gustavson2000, Peters2001, Canuel2006}. The following development of cold atom techniques lead to a large number of compact, stable and accurate experiments investigating both fundamental and applied aspects. Indeed, atom interferometry rapidly found a large range of applications such as measurement of fundamental constants \cite{Fixler2007, Rosi2014, Bouchendira2011, Parker191}, gravimetry \cite{Peters99,Gillot2014,Freier2016}, gradiometry \cite{Snadden98}, underground survey \cite{deAngelis2009, Metje2011} and inertial navigation \cite{KasevichPatent,Battelier2016}. The relentless progress in terms of sensitivity of AI based instruments now enables to investigate fundamental questions like those related to the interface of gravity and quantum mechanics \cite{Asenbaum2017,PhysRevLett.120.043602}. The ability of AI to reveal extremely small changes in the inertial field can be used for General Relativity (GR) tests such as the validation of the Weak Equivalence Principle \cite{Dimopoulos2007,Geiger2011,Schlippert2014,Barrett2015,Barrett2016,Rosi2017} and the local Lorentz invariance of post-Newtonian gravity \cite{Muller2008}. Such experiments would particularly profit from micro gravity environments allowing for extremely long interaction times, and therefore several GR space missions based on AI were also proposed to measure for example the Lens-Thirring effect \cite{Jentsch2004} and the Weak Equivalence Principle \cite{Aguilera2014, Schubert2013, Tino2013}. On the base of these developments, the idea also emerged to use AI for the detection of Gravitational Waves (GWs) at low frequency \cite{Borde1983,Chiao2004,Foffa2006,Tino2007}.

The first direct GW detection by the two giant optical interferometers of Advanced LIGO \cite{GWdet} in September 2015 opens a new area for physics: in the future, GW detectors will reveal new information about massive astrophysical objects such as neutron stars, black holes, pulsars and their dynamics. The transient signal of this first observation, a chirp in frequency from 35 to 250 Hz lasting about 150 ms with a peak strain amplitude of 10$^{-21}$, corresponds to the merging phase of a black hole binary system. Due to their bandwidth limited to the frequency range 10 Hz-10 kHz, only the last evolution phase of binary systems is observable with current GW detectors. Before their coalescence, the same sources emit at lower frequencies quasi-continuous GW signals in their ``inspiral'' phase. A new class of low frequency detectors would enable to observe the signal of such sources years before they enter in the bandwidth of ground-based optical detectors. For example, this first observed source, GW150914, was emitting at a frequency of 16 mHz 5 years before coalescence with a characteristic strain amplitude of the order of 10$^{-20}$ \cite{Sesana}. Low frequency GW detectors would therefore open the possibility of multi-band GW astronomy with long term observation of all evolution phases of binary systems \cite{Sesana}. Such observatories would also enable to precisely predict the event of coalescence in time and space \cite{Graham2018}, which would ease coincident observations in the electromagnetic domain \cite{Abbott2017}. Multi-band GW astronomy then offers a great scientific payoff with the perspective of multi-messenger astronomy \cite{Christensen}, but also promises new gravity and cosmology tests. Low frequency detectors would therefore open a completely new area for GW astronomy \cite{Mandel2018}. In this context, an original concept for GW detection appeared that combines AI methods and laser ranging techniques used in optical GW detectors. It consists in measuring by AI the effect of GWs on a laser link correlating distant atomic sensors \cite{Dimopoulos2009,Dimopoulos2008,Harms2013}. Such approach promises to overcome many limitations that affect purely optical interferometers and opens the way toward sub-Hz GW observation.

This new application gave birth to a new field, \textit{large scale atom interferometry}. The use of AI techniques for the realization of large instruments will drastically increase applications of matter-wave interferometry by offering measurements of space-time fluctuations of the gravitational forces along one direction of space. Such instruments will open groundbreaking perspectives not only for GW detection but also for geosciences, allowing for precise mapping of mass distributions and transfers around the detector. This field will be explored by the Matter wave-laser based Interferometer Gravitation Antenna (MIGA) \cite{Canuel2014, Canuel2016}, the first very long baseline AI instrument, now under construction. This detector will consist of a network of AI sensors interrogated by the resonant field of an optical cavity, a configuration allowing for an interferometric control of the phase front of the manipulation beams \cite{Hamilton2015}, as well as an increase of the instrument sensitivity thanks to the high momentum transferred to the atomic wavefunction \cite{Riou2017}. The instrument will be located in dedicated tunnels, 500 m underground at the ``Laboratoire Souterrain {\`a} Bas Bruit'' (LSBB), in Rustrel (France) in an environment with very low background noise and located away from major anthropogenic disturbances. MIGA will offer a multidisciplinary approach for the study of both fundamental and applied aspects of gravitation by correlating the main detector's signal with highly-precise instrumentation specific to geoscience.
In this article, we will detail the measurement strategy of MIGA, present its general design and analyze the noise contributions and expected sensitivity, in relation to both strain measurements and detection of geophysical phenomena.

\color{black} 
\section{MIGA measurement scheme}
 
\subsection{Combining atom and laser interferometry}

\begin{figure}[htp]
\centering
\includegraphics[width=15cm]{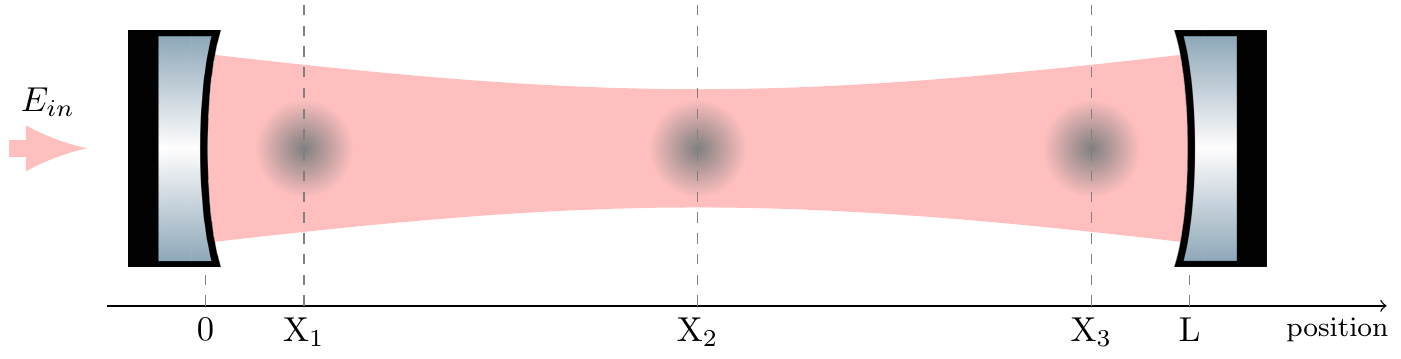}
\caption{Scheme of the MIGA antenna. A set of three atom interferometers at positions X$_{1,2,3}$ are interrogated by the resonant field of an optical cavity.}
\label{MigaScheme}
\end{figure}

The working principle of MIGA consists in correlating a set of three distant light pulse atom interferometers using the resonant field of a long ultra-stable optical cavity (see Fig. \ref{MigaScheme}). 

Simultaneous matter wave interferometers along the cavity are realized by time-modulation of the laser injected inside the resonator. The signal at the output of each interferometer will depend on the relative phase accumulated along the different interferometric path followed by the matter waves. The atom interferometric phase is linked to the phase of the cavity field, imprinted on the matter waves during each interrogation pulse. 

The response of each atom interferometer therefore depends on local \textbf{inertial forces} sensed by the atoms, or any other effects modifying the optical phase seen by the matter waves during the interferometer sequence. Such effects may arise from \textbf{strain variation} inside the optical cavity induced by the effect of GWs or some residual motion of the cavity mirrors. MIGA is therefore an hybrid atom-laser interferometer using an array of atom sensors that will simultaneously measure motion of the cavity, GWs and inertial effects.

\subsection{Geometry of MIGA atom interferometers}

\begin{figure}[htp]
\centering
\includegraphics[width=10cm]{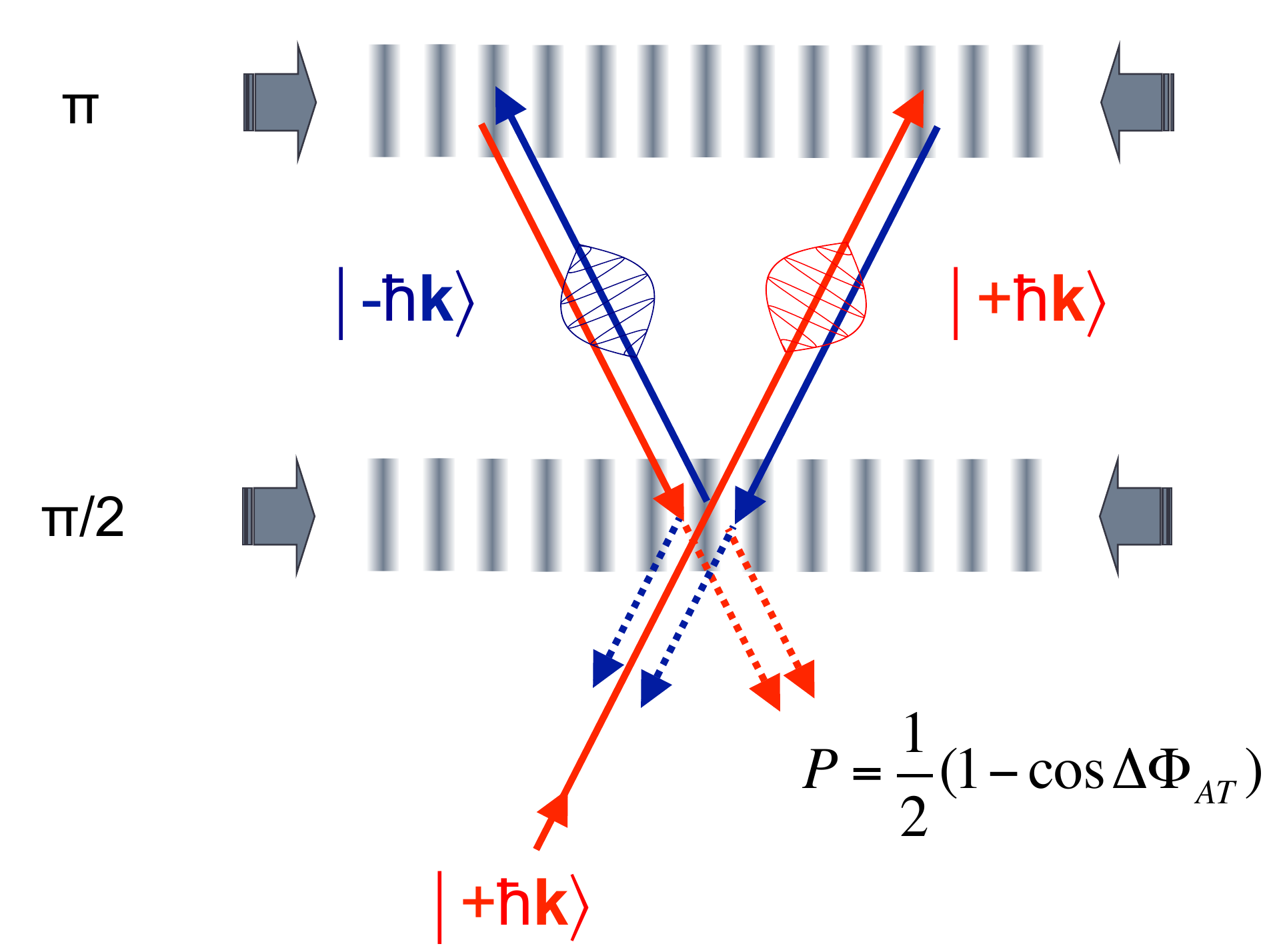}
\caption{Principle of a Mach Zehnder interferometer for matter waves.}
\label{MachZ2}
\end{figure}

MIGA will make use of 3 pulses Mach-Zehnder Interferometers, interrogated in the Bragg regime by two horizontal cavity fields (see Fig. \ref{MachZ2}). 

After launching on a vertical trajectory, the atoms experience a set of  $\pi/2$-$\pi$-$\pi/2$ cavity pulses that separate, deflect and recombine the matter waves. The energy-momentum conservation during the light-matter interaction implies to couple atomic states of momenta $\left |+n\hbar \mathbf{k}\right\rangle$ and $\left |-n\hbar \mathbf{ k}\right\rangle$ where $\mathbf{k}$ is the wave vector of the interrogation field and $n$ the Bragg diffraction order. 
The transition probability $P$ between the different states at the interferometer output is given by a two wave interference formula $P=1/2 \left[1-\cos (\Delta \phi_{AT}) \right]$ where $\Delta \phi_{AT}$ is the atom phase shift accumulated inside the interferometer.

Using the sensitivity function formalism \cite{Cheinet08}, the response of the atom interferometer at time $t$ can be expressed as a function of the variations of the local phase difference $\Delta \varphi (t)$ for the two counter-propagating interrogation fields:
\begin{equation}
\Delta \phi_{AT}(t)=n \int^{\infty}_{-\infty}s(T-t)\frac{d \, \Delta \varphi (T)}{dT}dT= n \, \Delta \varphi(t) \ast \frac{d s(t)}{dt}
\label{AIresponse}
\end{equation}
where $s(t)$ is the sensitivity function of a three pulse atom interferometer, detailed in \cite{Cheinet08}, and "$\ast$" denotes the convolution.

\subsection{\label{par:cavity}Response of an in-cavity atom interferometer}

\begin{figure}[!h]
	\centering
\includegraphics[width=15cm]{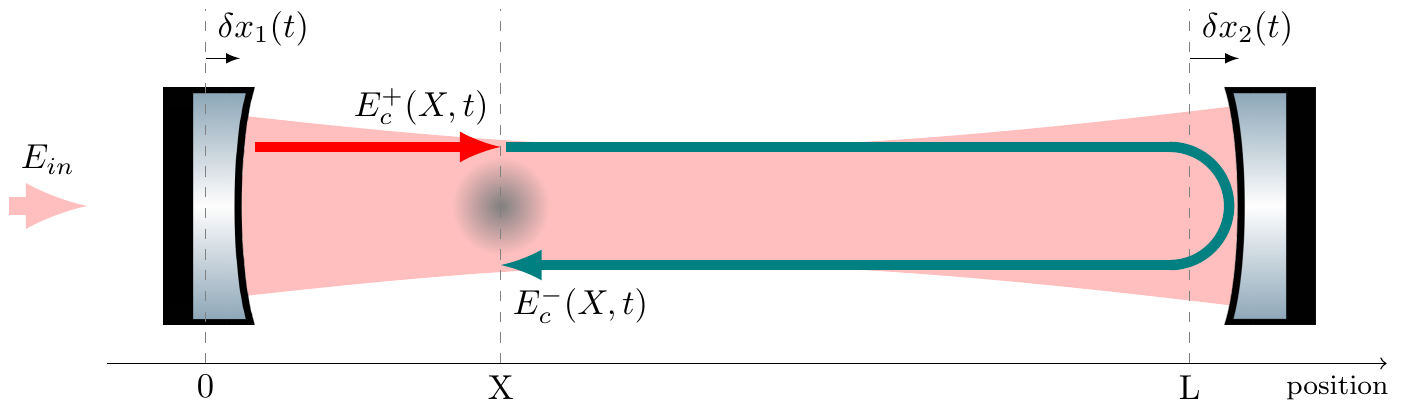}
\caption{An atom source placed at position $X$ along the cavity is interrogated by the counter-propagating fields $E_c^{+}(X,t)$ and $E_c^{-}(X,t)$. $\delta x_1(t)$ and $\delta x_2(t)$ are respectively the position fluctuations of the left and right mirrors with respect to the cavity baseline at rest $L$.}
\label{MIGAScheme2}
\end{figure}

In the following, we derive the response of an atom interferometer interrogated by a cavity at resonance in presence of strain variations induced by GWs, taking into account the major noise contributions, due to cavity mirror vibrations and frequency noise of the input optical field. We will consider that the GW propagates in the direction perpendicular to the cavity axis $x$. The notations used for the calculation are summarized in Fig.~\ref{MIGAScheme2}. We describe the electromagnetic field in the cavity as a superposition of two counter-propagating waves, $E_c^{\pm}(t)$, respectively propagating towards positive and negative $x$. As stated by Eq. \ref{AIresponse}, the atom interferometer response is determined by the \textit{relative} phase $\Delta \varphi (t)$ between these fields, which is imprinted on the atoms at position $X$. We will determine $\Delta \varphi (t)$ as a function of monochromatic variations of the mirror positions $\delta x_{1,2}(t)=$ 
$\delta x_{1,2}\cos (\Omega_{x_{1,2}}t)$, the frequency noise of the input laser $\delta\nu_{in}(t)= 
\delta \nu_{in}\cos (\Omega_{\nu_{in}}t)$, and the GW strain amplitude $h(t)=h\cos (\Omega_{gw}t)$.
%

At steady state, the intra-cavity field propagating in the forward direction at position X can be written in the general form:
\begin{equation*}
E_c^+(t) = E_{c0}(t) e^{i\varphi^+(t)}.
\end{equation*}
At time t, photons from the field propagating in the backwards direction at position X, $E_c^-(t)$ are emitted from the forward field at retarded time $t_r=t-\Delta t_r$, where $\Delta t_r$ is the light propagation delay on the round trip between the AI and the cavity end mirror. We then have:
\begin{equation*}
E_c^-(t)= E_c^+(t_r)=E_c^+(t-\Delta t_r)
\end{equation*}
which means:
\begin{equation*}
\varphi^-(t)=\varphi^+(t-\Delta t_r).
\label{}
\end{equation*}
To first order in $\Delta t_r$, considering that the phase variations are slow with respect to the cavity round trip, one obtains:
\begin{equation*}
\varphi^-(t)=\varphi^+(t)-\Delta t_r \frac{d \varphi^+(t)}{dt}
\label{}
\end{equation*}
\begin{equation}
\Delta \varphi (t)=\Delta t_r \frac{d \varphi^+(t)}{dt}=2\pi\Delta t_r \nu (t).
\label{EQUdphi}
\end{equation}
The relative phase imprinted on the atom can be expressed as the product of the propagation delay and the instantaneous frequency of the intracavity field at steady state $\nu (t)$; we calculate both terms thereafter.

\subsection*{Calculation of $\nu (t)$}
The frequency of the intracavity field $\nu(t)$ can be expressed as the sum of the different contributions:
\begin{equation}
\nu (t)=\nu_0 +\delta\nu_{cin}(t) + \delta \nu_{\delta x_1}(t)+ \delta \nu_{\delta x_2}(t)+ \delta \nu_{gw}(t)
\end{equation}
where $\delta\nu_{cin}(t)$, $\delta \nu_{\delta x_{1,2}}(t)$ and $\delta \nu_{gw}(t)$ accounts for fluctuations of the intra-cavity field frequency induced by variations of the input frequency, vibration of cavity mirrors and GW effects. 
Calculation of $\delta \nu_{\delta x_{1,2}}(t)$ and $\delta \nu_{gw}(t)$ (to first order in $\frac{\Omega_{x_{1,2}}}{\omega_p }$ and $\frac{\Omega_{gw}}{\omega_p }$) are carried out in annex \ref{AnnexA} and \ref{AnnexB}:
\begin{align*}
\delta\nu_{gw}(t)&=\frac{1}{2}\frac{\Omega_{gw}}{\omega_p }\nu_0 h \sin (\Omega_{gw}t)\\
\delta\nu_{\delta x_{1,2}}(t)&=\pm\frac{\Omega_{x_{1,2}}}{\omega_p }\nu_0\frac{\delta x_{1,2}}{L} \sin (\Omega_{x_{1,2}}t)
\end{align*}
where $\omega_p/2\pi=c/(4LF)$ is the frequency pole of the cavity, and $F$ its finesse. Since the cavity acts as a first order low pass filter for fluctuations of the input frequency \cite{Zhu1993}, we have:
\begin{equation*}
\delta\nu_{cin}(t)=\frac{\delta\nu_{in}\cos \left[\Omega_{\nu_{in}}t+\arctan\left(\frac{\Omega_{\nu_{in}}}{\omega_p}\right)\right]}{\sqrt{1+\left(\frac{\Omega_{\nu_{in}}}{\omega_p}\right)^2}}.
\end{equation*}
To first order in $\frac{\Omega_{\nu_{in}}}{\omega_p}$ we have therefore: 
\begin{equation*}
\delta\nu_{cin}(t) \simeq \delta\nu_{in}\cos \left(\Omega_{\nu_{in}}t\right)-\frac{\Omega_{\nu_{in}}}{\omega_p}\delta\nu_{in}\sin \left(\Omega_{\nu_{in}}t\right).
\end{equation*}
The instantaneous frequency of the intracavity field can then be expressed:
\begin{multline}
\nu (t)=\nu_0\left(1 +\frac{\delta\nu_{in}}{\nu_0}\cos \left(\Omega_{\nu_{in}}t\right)-\frac{\Omega_{\nu_{in}}}{\omega_p}\frac{\delta\nu_{in}}{\nu_0}\sin \left(\Omega_{\nu_{in}}t\right)+\right.\\ 
\left.\frac{\Omega_{x_{1}}}{\omega_p }\frac{\delta x_{1}}{L} \sin (\Omega_{x_{1}}t)-\frac{\Omega_{x_{2}}}{\omega_p }\frac{\delta x_{2}}{L} \sin (\Omega_{x_{2}}t)+\frac{1}{2}\frac{\Omega_{gw}}{\omega_p }h \sin (\Omega_{gw}t)\right).
\end{multline}

\subsection*{Calculation of the propagation delay $\Delta t_r$}
The relativistic invariant for the electromagnetic field propagating in the $x$ direction can be written:
\begin{equation*}
ds^2 = c^2dt^2-dx^2 + h(t)dx^2 =0. 
\end{equation*}
To first order in $h$: 
\begin{equation}
dx = \pm [1+\frac{1}{2} h(t)] c dt
\label{eq:invar}
\end{equation}
where the plus (resp. minus) sign corresponds to the light propagating along the cavity from left to right (resp. right to left).
Photons from the resonant field $E_c^{+}$ are emitted at position $X$ at retarded time $t_r$ and then reflected on the cavity end mirror at position $L + \delta x_2(t_1)$ at time $t_{r'}$. Integrating Eq. \ref{eq:invar} brings to:
\begin{equation}
(L-X) + \delta x_2(t_{r'}) = c\left( t_{r'}-t_r \right) +\frac{c}{2} \int_{t_r}^{t_{r'}} h(\xi)d\xi.
\label{eq:t11}
\end{equation}
Photons are then reflected back and return to position $X$ at time $t$:
\begin{equation}
-(L-X)-\delta x_2(t_{r'}) = c \left( -t+t_{r'} \right) -\frac{c}{2} \int_{t_{r'}}^{t} h(\xi)d\xi.
\label{eq:t12}
\end{equation}
Subtracting Eq. \ref{eq:t11} and \ref{eq:t12}, the emission time $t_r$ of the photon can be expressed:
\begin{equation}
t_r = t -\frac{2(L-X)}{c}-\frac{2\delta x_2(t_{r'})}{c}+\frac{1}{2}\int_{t_r}^{t} h(\xi)d\xi. 
\end{equation}
This implicit equation can be simplified to first order in $h$ and $\delta x_2$:
\begin{equation}
t_r = t -\frac{2(L-X)}{c}-\frac{2\delta x_2(t-\frac{(L-X)}{c})}{c}+\frac{1}{2}\int_{t -\frac{2(L-X)}{c}}^{t} h(\xi)d\xi. 
\label{eq:RetartedTime}
\end{equation}
$\Delta t_r= t- t_r$ can then be expressed as:
\begin{equation}
\Delta t_r=\frac{2(L-X)}{c}+\frac{2\delta x_2(t-\frac{(L-X)}{c})}{c}-h\frac{(L-X)}{c}\textrm{sinc} \left(\Omega_{gw}\frac{(L-X)}{c}\right)  \cos \left(\Omega_{gw}\left(t-\frac{(L-X)}{c}\right)\right).
\end{equation}
Expanding in series the cosine terms, we obtain to first order in $\Omega_{gw}L/c$ and $\Omega_{x_{2}}L/c$:

\begin{multline}
\Delta t_r=\frac{2(L-X)}{c}+\frac{2\delta x_2}{c}\left( \cos (\Omega_{x_{2}}t) +\Omega_{x_{2}}\frac{(L-X)}{c}\sin (\Omega_{x_{2}}t)\right)\\
-h\frac{(L-X)}{c} \left( \cos (\Omega_{gw}t) +\Omega_{gw}\frac{(L-X)}{c}\sin (\Omega_{gw}t)\right).
\end{multline}

\subsection*{Calculation of $\Delta \phi(t)$}
From Eq. \ref{EQUdphi}, by keeping only first order terms in $\delta\nu$,$\delta x_{1,2}$ and $h$:
%
\begin{multline*}
  \Delta \varphi (t)=4\pi\nu_0\frac{(L-X)}{c} \left(1+\frac{\delta\nu_{in}}{\nu_0}\cos \left(\Omega_{\nu_{in}}t\right)-\frac{\Omega_{\nu_{in}}}{\omega_p}\frac{\delta\nu_{in}}{\nu_0}\sin \left(\Omega_{\nu_{in}}t\right)-\frac{h}{2}\cos (\Omega_{gw}t)\right.\\
  \left.-\frac{h}{2}\Omega_{gw}\left[\frac{L-X}{c}-\frac{1}{\omega_p}\right] \sin (\Omega_{gw}t)+\frac{\delta x_1}{L}\frac{\Omega_{x_{1}}}{\omega_p}\sin (\Omega_{x_{1}}t)+\frac{\delta x_2}{L}\Omega_{x_{2}}\left[\frac{L}{c}-\frac{1}{\omega_p}\right]\sin (\Omega_{x_{2}}t)\right)\\
  +\frac{4\pi\nu_0}{c}\delta x_2 \cos (\Omega_{x_{2}}t).
\end{multline*}
The terms $\frac{1}{\omega_p}$ and $\frac{L}{c}$ are respectively the cavity photon lifetime and half of the cavity transit time; for $F\gg1$ it holds $\left[\frac{L}{c}-\frac{1}{\omega_p}\right]\approx-\frac{1}{\omega_p}$. We thus obtain: 
\begin{multline*}
  \Delta \varphi (t)=4\pi\nu_0\frac{(L-X)}{c} \left(1+\frac{\delta\nu_{in}}{\nu_0}\cos \left(\Omega_{\nu_{in}}t\right)-\frac{h}{2}\cos (\Omega_{gw}t)+\frac{1}{\omega_p}\left[-\Omega_{\nu_{in}}\frac{\delta\nu_{in}}{\nu_0}\sin \left(\Omega_{\nu_{in}}t\right)\right.\right.\\
  \left.\left.+\frac{h}{2}\Omega_{gw} \sin (\Omega_{gw}t)+\frac{\delta x_1}{L}\Omega_{x_{1}}\sin (\Omega_{x_{1}}t)-\frac{\delta x_2}{L}\Omega_{x_{2}}\sin (\Omega_{x_{2}}t)\right]\right)\\+\frac{4\pi\nu_0}{c}\delta x_2 \cos (\Omega_{x_{2}}t).
\end{multline*}
Considering time-fluctuating effects with characteristic frequencies smaller than the frequency pole of cavity, we obtain:
\begin{multline}
  \Delta \varphi (t)=4\pi\nu_0\frac{(L-X)}{c} \left(1+\frac{\delta\nu(t)}{\nu_0}-\frac{h(t)}{2}+\frac{1}{\omega_p}\left[\frac{\delta\nu^{'}(t)}{\nu_0}-\frac{h^{'}(t)}{2}+\frac{\delta x^{'}_2(t)-\delta x^{'}_1(t)}{L}\right]
  \right)\\
  +\frac{4\pi\nu_0}{c} \delta x_2(t)
\end{multline}
where character "$'$" denotes the time derivative.
%
The response of the AI can be expressed from Eq. \ref{AIresponse}:
\begin{equation}
\Delta\phi_{AT}(X,t)=n\Delta \varphi (X,t) \ast s^{'}(t)+\Delta\phi_{I}(X,t)+\alpha(X,t).
\label{AIresponse2}
\end{equation}
In this expression, we introduced the detection noise $\alpha(X,t)$ and the inertial signal $\Delta\phi_{I}(X,t)$ associated with the local gravitational acceleration experienced by the atoms. The latter term reads:
\begin{equation}
\Delta\phi_I(X,t)=\frac{4n\pi\nu_0}{c} x(X,t) \ast s^{'}(t)
\end{equation}
where $x(X,t)$ represents the motion of the atoms along the laser beam direction due to the fluctuations of the local gravity.

%
The differential signal between two atom interferometers placed in cavity at position $X_1$ and $X_2$ is therefore:
\begin{multline}\label{AIresponse3}
\Delta\phi_{AT}(X_1,t)-\Delta\phi_{AT}(X_2,t)=\frac{4n\pi\nu_0}{c}(X_2-X_1)\left(\frac{\delta\nu(t)}{\nu_0}-\frac{h(t)}{2}+\frac{x(X_1,t)-x(X_2,t)}{X_2-X_1}
  \right.\\
  \left.+\frac{1}{\omega_p}\left[\frac{\delta\nu^{'}(t)}{\nu_0}-\frac{h^{'}(t)}{2}+\frac{\delta x^{'}_2(t)- \delta x^{'}_1(t)}{L}\right]  \right) \ast s^{'}(t)+\alpha(X_1,t)-\alpha(X_2,t).
\end{multline}
Eq. \ref{AIresponse3} shows that an in-cavity gradiometer has a response similar to a free space one, with additional terms in $\frac{1}{\omega_p}$ taking into account the response of the cavity to the different noises sources. Moreover, in analogy with the free space configuration, the in-cavity gradiometer presents a strong reduction of the influence of end cavity mirror vibrations with respect to the signal of a single AI.

\section{Gravitational Waves physics with MIGA}\label{migaGWphysics}

\subsection{Strain sensitivity to GW}\label{GWstrainEq}
In this section we derive the sensitivity of the MIGA instrument to GW strain variation. For sake of simplicity, we will only  consider the signal $\Gamma(t)$ obtained from the largest atom gradiometer available from the antenna that will use atom sources close to the cavity mirors (i.e. $X_1=0$ and $X_2=L$) :
\begin{equation}
\Gamma(t)=\Delta\phi_{AT}(0,t)-\Delta\phi_{AT}(L,t).
\label{AIresponse2}
\end{equation}
We calculate the strain sensitivity that can be obtained from this configuration, defined as the minimum detectable GW Power Spectral Density (PSD).
Considering that the fluctuations $\delta\nu(t)$, $h(t)$, $\delta x_1(t)$, $\delta x_2(t)$, $\alpha(X_1,t)$, $\alpha(X_2,t)$ and the strain Newtonian Noise $\mathrm{NN}(t)=\frac{x(0,t)-x(L,t)}{L}$ are uncorrelated, the PSD $S_{\Gamma}$ of the gradiometer signal can be expressed:

\begin{multline}
S_{\Gamma}(\omega)=\left[\frac{4n\pi\nu_0}{c}L\right]^2\left(\left[1+\frac{\omega^2}{\omega_p^2}\right]\left(\frac{S_{\delta\nu}(\omega)}{\nu_0^2}+\frac{S_h(\omega)}{4}\right)+\frac{2\omega^2S_x(\omega)}{\omega_p^2L^2}+S_{\mathrm{NN}}(\omega)\right)|\omega G(\omega)|^2\\
  +2S_{\alpha}(\omega)
\label{SignalPSD}
\end{multline}
where $G(\omega)$ is the Fourier transform of the AI sensitivity function $s(t)$, and $S_{u}(\omega)$ denotes the PSD of the fluctuations of $u(t)$. We consider in Eq. \ref{SignalPSD} that $S_{x_1}(\omega)=S_{x_2}(\omega)=S_{x}(\omega)$ and $S_{\alpha(X_1,.)}(\omega)=S_{\alpha(X_2,.)}(\omega)=S_{\alpha}(\omega)$.

Using the gradiometer signal, the signal to noise ratio $\mathrm{SNR}(\omega)$ for a GW detection at a frequency $\omega$ is defined by the ratio between the GW term and all other terms of Eq. \ref{SignalPSD}:
\begin{equation}
\mathrm{SNR}(\omega)=\frac{\left[1+\frac{\omega^2}{\omega_p^2}\right]\frac{S_h(\omega)}{4}}{\left[1+\frac{\omega^2}{\omega_p^2}\right]\frac{S_{\delta\nu}(\omega)}{\nu_0^2}+\frac{2\omega^2S_x(\omega)}{\omega_p^2L^2}+S_{\mathrm{NN}}(\omega)+\frac{2S_{\alpha}(\omega)}{\left[\frac{4n\pi\nu_0}{c}L\right]^2|\omega G(\omega)|^2}}.
\label{AIresponse2}
\end{equation}
The strain sensitivity to GW is then defined by the GW PSD which corresponds to a SNR of 1:
\begin{equation}
S_h(\omega)=\frac{4S_{\delta\nu}(\omega)}{\nu_0^2}+\frac{4}{\left[1+\frac{\omega^2}{\omega_p^2}\right]}\left(\frac{2\omega^2S_x(\omega)}{\omega_p^2L^2}+S_{\mathrm{NN}}(\omega)+\frac{2S_{\alpha}(\omega)}{\left[\frac{4n\pi\nu_0}{c}L\right]^2|\omega G(\omega)|^2}\right).
\label{sensitivity}
\end{equation}

\subsection{MIGA strain sensitivity curve and influence of instrumental noise}\label{GWstrain}
We now determine the strain sensitivity curve of MIGA from the projection of the different noise sources identified in Eq. \ref{sensitivity}. The antenna will use a resonator of length $L=200$ m with an initial Finesse of $F=100$ that corresponds to a cavity frequency pole of $\frac{\omega_p}{2\pi}= 2.5\ $~kHz (see Sec. \ref{cavityInterro} for a discussion on cavity parameters). We will consider the use of Rb atom sources interrogated on the $D_2$ line, setting the wavelength of the resonating field to $\frac{c}{\nu_0}= 780\ $~nm. The total interrogation time of the atom will be $2T=500$ ms and the Rabi frequency $\frac{\Omega_{eff}}{2\pi}= 10\ $~kHz.

\begin{figure} [h!]
   \begin{center}
   \begin{tabular}{c} 
   \subfigure
   {{\includegraphics[width=.6\paperwidth,trim={0cm 0 1.5cm 4cm},clip]{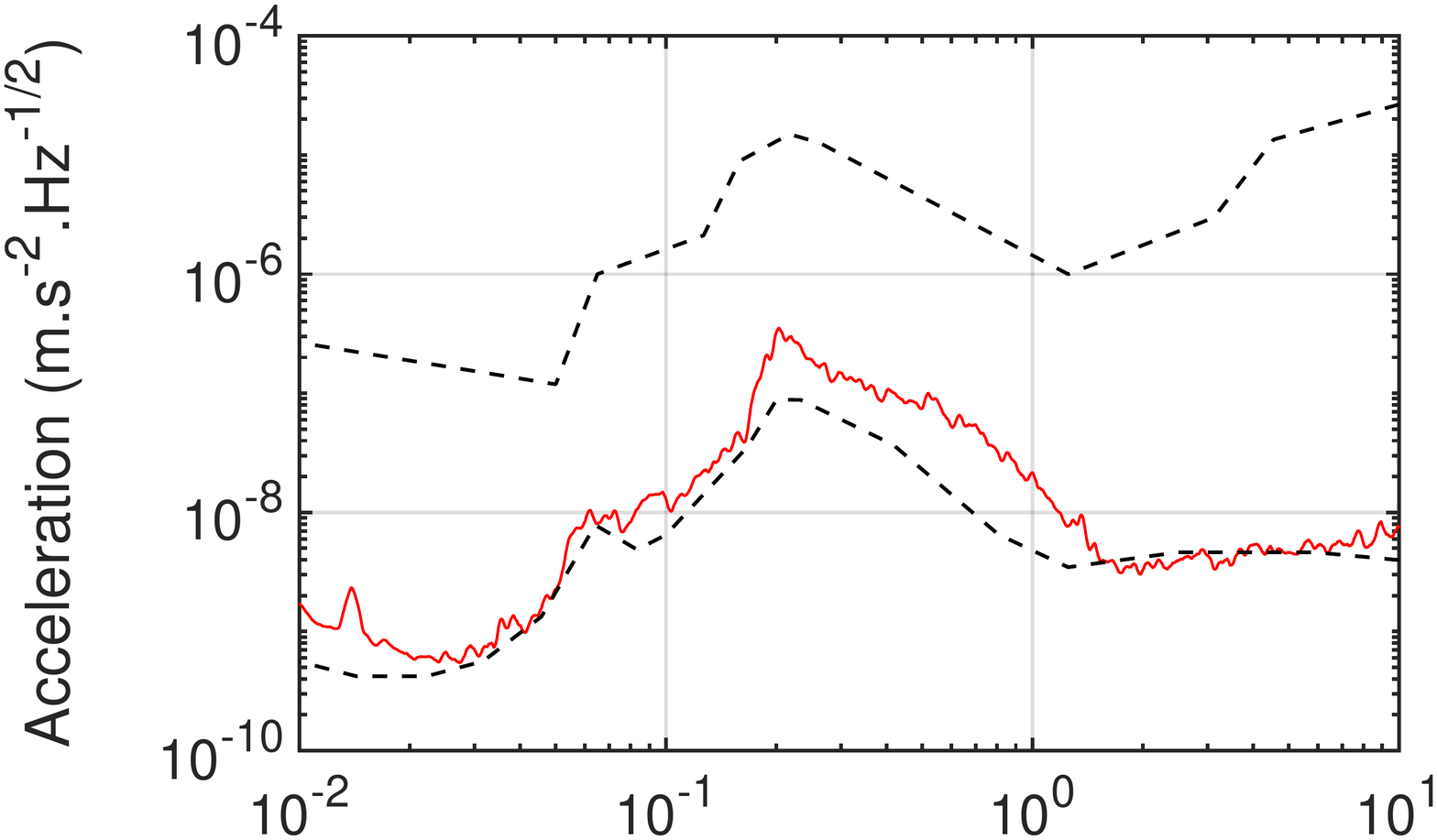}}}\\
    \subfigure{{\includegraphics[width=.6\paperwidth,trim={0 0 1.5cm 0cm},clip]{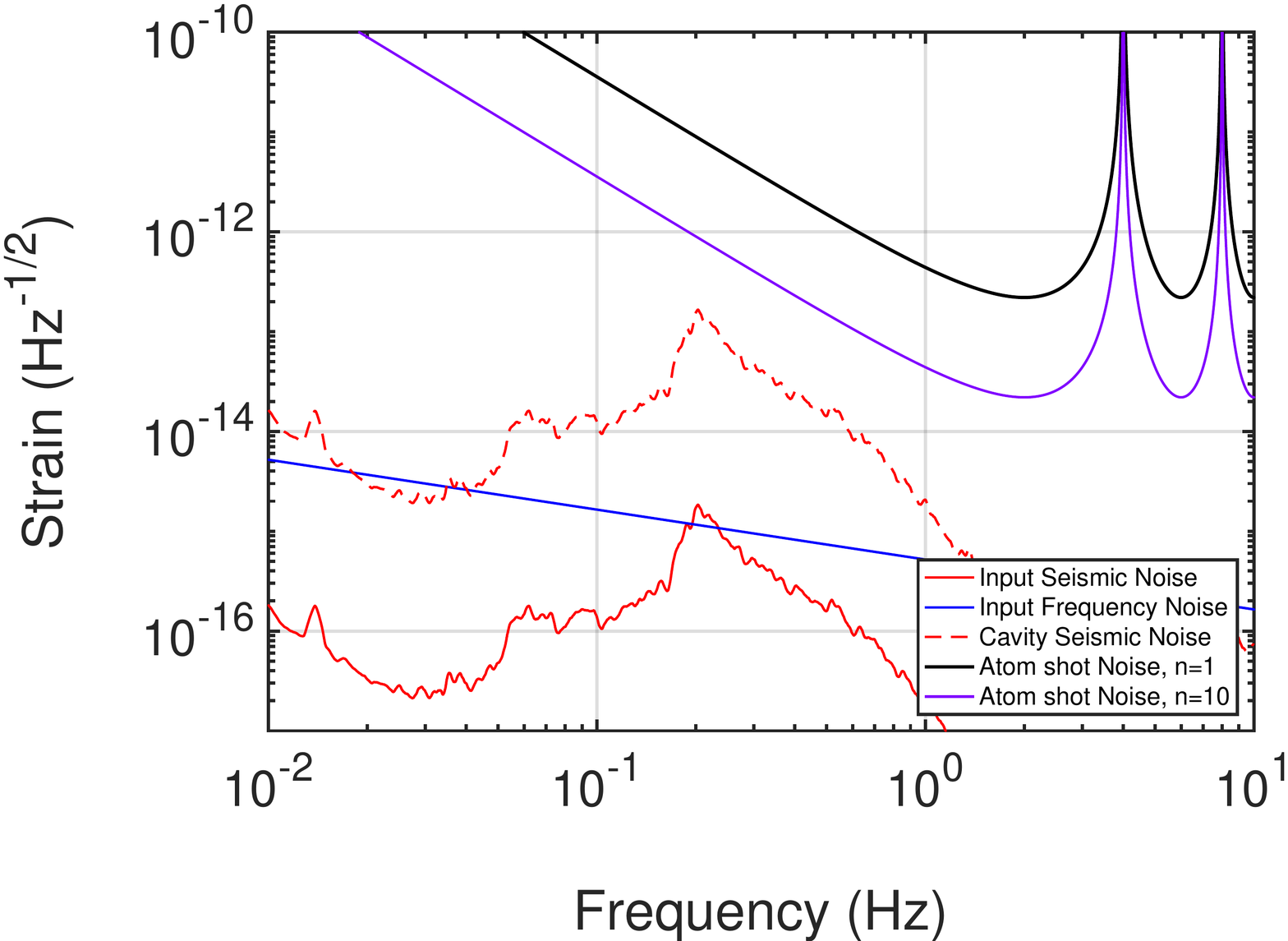}}}
   \end{tabular}
   \end{center}
   \caption[example] 
   {\label{SensitivityCurve}Top: average seismic acceleration PSD measured in the LSBB galleries with a STS-2 sensor during a typical quiet period. Bottom: projection of the different noise sources on strain sensitivity.}
\end{figure}

We consider that the laser at the input of the interrogation cavity will be pre-stabilized with a state-of-the-art reference cavity. The relative frequency noise of the input laser becomes thus limited by the thermal noise of the ultra-stable reference cavity at the level of $S_{\delta\nu}(f)=0.01\times (1\ \mathrm{Hz}/f)\mathrm{Hz}^2/\mathrm{Hz}$ \cite{Numata2004,Thorpe2011,Jiang2011}. The projection of such noise on strain sensitivity (first term of Eq.~\ref{sensitivity}) is plotted in blue on Fig.~\ref{SensitivityCurve}-bottom.

The Newtonian Noise is a gravity noise sensed by the atom interferometers of the antenna due to density fluctuations of the medium surrounding the detector. Such noise was extensively studied in the field of optical GW detectors \cite{Saulson,Harms2015} and the two main sources at low frequencies are coming from seismic and atmospheric perturbations. The third term of Eq.~\ref{sensitivity} represents the sum of these contributions for the atom sources of MIGA. These contributions were previously studied in \cite{Canuel2016} at LSBB, showing a limit of the strain sensitivity smaller than 10$^{-16}$ Hz$^{-1/2}$ above 0.1 Hz.

The displacement noise $S_x(\omega)$ of the cavity optics that will affect the measurements will depend on the system used for suspending and controlling the cavity. To obtain an upper limit of such noise, we consider the worst case of non-suspended mirrors. Fig.~\ref{SensitivityCurve}-top shows the average seismic acceleration PSD measured in the LSBB galleries with a STS-2 sensor during a typical quiet period (see section \ref{Par:lsbb_lowNoise} for a discussion on seismic properties at the LSBB laboratory). According to Eq.~\ref{sensitivity}, displacement noise of cavity mirrors can contribute to limit strain sensitivity through phase modulation of the intra-cavity field (second term of Eq.~\ref{sensitivity}). This contribution is plotted in dashed red in Fig.~\ref{SensitivityCurve}-bottom. 

It must be noted that seismic noise may also affect input optics and create an extra frequency noise on the input laser $S_{\delta\nu}'(\omega)=\frac{\nu_0^2}{c^2}\omega^2S_{xin}(\omega)$ \cite{Chaibi2016}, where $S_{xin}$ is the displacement noise PSD of the input optics. The projection of this extra input laser frequency noise is plotted in solid red in Fig.~\ref{SensitivityCurve}-bottom. 

The last limitation to strain sensitivity in Eq.~\ref{sensitivity} comes from detection noise. At the output of each AI, the atom phase is measured by the fluorescence of the atom clouds. The noise in such process comes from the randomness associated with the quantum projection process, and will be $S_{\alpha}(\omega)=1\ (\mathrm{mrad})^2/\mathrm{Hz}$ for an atom flux of $10^6$ atom/s. The projection of detection noise on strain measurement is plotted in Fig.~\ref{SensitivityCurve}-bottom considering Bragg transition of order $n=1$ and $n=10$ (respectively solid black and solid violet curves).

We observe that in the initial instrument configuration corresponding to $n=1$, the detection noise will be largely dominant on strain measurement in all detector bandwidth, setting an optimum strain sensitivity of 2.2$\cdot 10^{-13}$ at frequency $1/(2T)=2\ \mathrm{Hz}$.

\color{black}

\color{black}
\section{MIGA antenna design}

For the realization of the atom--laser antenna we rely on robust and well tested technology for the atom interferometers:

\begin{itemize}
\item rubidium atoms, the workhorse solution in AI; 
\item laser cooled atomic sources; 
\item free falling atom sensors;
\item two photon Bragg transitions for the coherent manipulation of the matter waves;
\item quantum-projection-noise limited sensitivity.
\end{itemize}

The unique exception to a standard atom interferometer is represented by the cavity enhanced interrogation (Par. \ref{par:cavity}), an essential component of the atom--laser antenna. For the design of the instrument to be installed underground at LSBB, we will adopt an interrogation time T=250 ms, and a cavity length of 200 m; these choices translate into an expected strain sensitivity of 2.2$\cdot$10$^{-13}$ at 2 Hz (Fig. \ref{SensitivityCurve}). Several solutions that go beyond this set of choices as a baseline device will be later taken into account to upgrade the instrument performance. The strain sensitivity of a gravity-gradiometer depends on the sensitivity to gravity acceleration of each atomic head, and increases linearly with the baseline size. Among the possibilities to tune the sensitivity of each atom interferometer we list the use of: long interrogation times, by adopting ultra-cold atomic sources \cite{Abend2016} or optically guided AI \cite{McDonald2013a}; large momentum splitting techniques \cite{McDonald2013,Asenbaum2017}; non-classical input states to achieve sub-shot-noise sensitivity \cite{Hosten2016,Cox2016}. Adopting alkali-earth-like atoms coherently manipulated on optical transitions could also mitigate the impact of the interrogation laser noise on long baseline interferometers \cite{Graham2013}.


\begin{figure}[!h]
\centering
\includegraphics[width=1.0\textwidth]{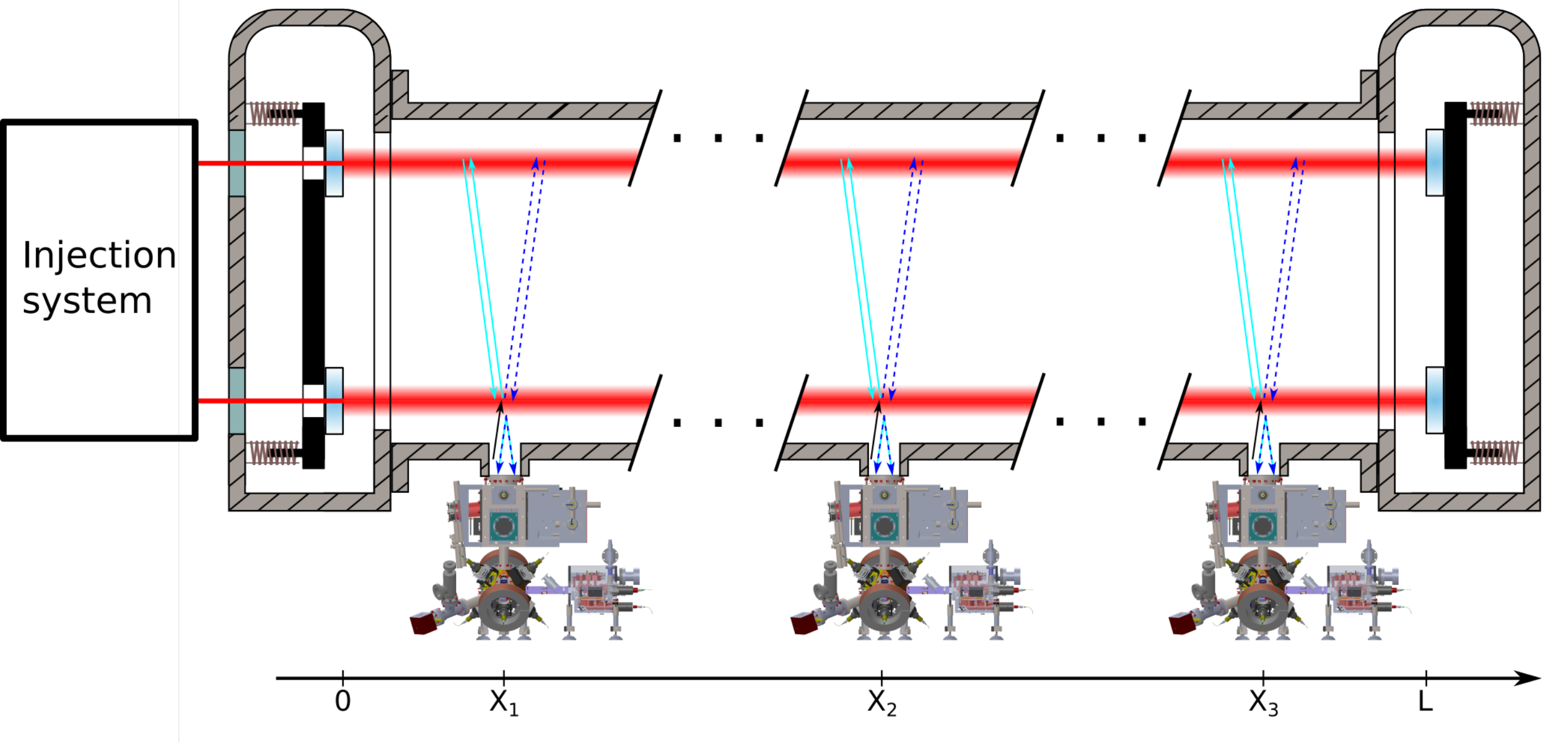}
\caption{Overview of the MIGA instrument with the main sub-systems. Three atomic heads at positions X$_{1,2,3}$ launch atomic clouds in an almost vertical parabolic flight (dotted lines); the atoms are manipulated in the upper part of the parabola with a Bragg interferometric sequence by way of light pulses at 780 nm (red horizontal lines) resonant with two horizontal cavities. The ultra-high-vacuum system encompassing the optical cavities, the mirrors payloads and their stabilization system is represented in gray; the atomic heads are connected to its lower side. The control system of the experimental setup, the laser systems dedicated to each atomic head, and the $\mu$-metal shield enclosing each interferometric region and the related atomic head are not represented in the figure.}
\label{fig:migaoverview}
\end{figure}

This size envisaged for the instrument brings AI from a laboratory scale to that of a big infrastructure, with consequent technological and engineering issues in terms of geometry, vacuum, magnetic field etc. In this section we will describe the experimental scheme of the MIGA detector, shown in Fig. \ref{fig:migaoverview}. Three atomic clouds are prepared at the same height and horizontally separated by a distance L, using for each a 3D-MOT loaded with a 2D-MOT. The atomic clouds are successively launched in free fall along the vertical direction using a moving optical molasses. Before entering in the interferometric region, the atomic ensembles are prepared in relation to their internal and external degrees of freedom, by means of two sets of horizontal Raman beams and optical pumping techniques. The horizontal acceleration at the position of the three atomic clouds is then simultaneously measured using Bragg interferometry, implemented with a $\pi/2$--$\pi$--$\pi/2$ pulse sequence symmetric with respect to the apogees of the atomic trajectories. After the Bragg pulse sequence, the atomic phase of each atom interferometer encoded as population difference of two atomic states is measured by fluorescence detection, when the clouds pass in the detection regions during their free fall.

\subsection{Atomic source}

\begin{figure}[!h]
\centering
\includegraphics[width=0.9\textwidth]{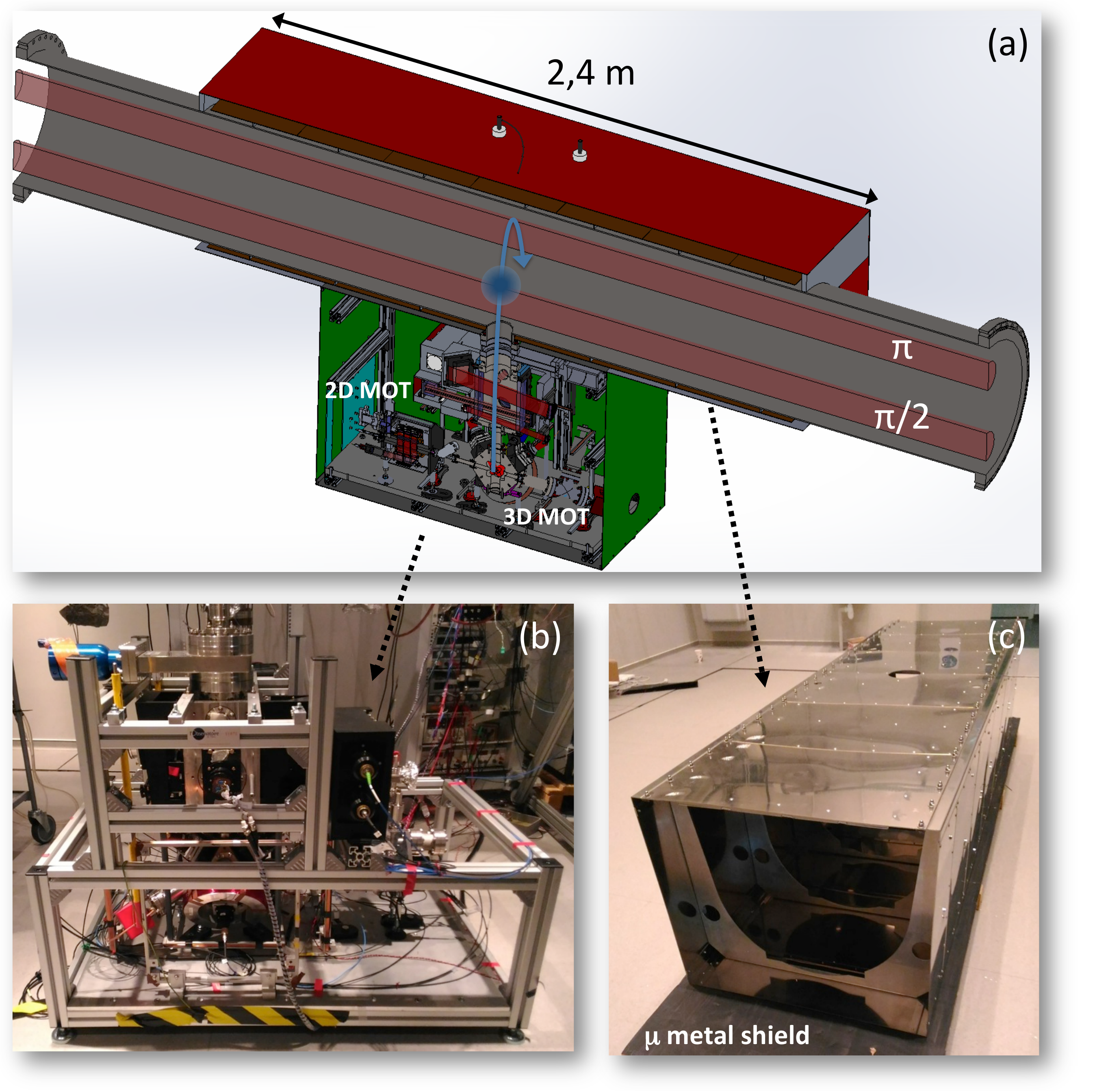}
\caption{Description of the MIGA atom interferometers, from \cite{Canuel2016}. (a) View of the AI including 2D-MOT, 3D-MOT, preparation and detection systems. A set of 3 different $\mu$-metal shields (red, brown, green colors) are used to screen the system from magnetic fields. (b) Picture of the 2D-MOT, 3D-MOT, preparation and detection systems. (c) Picture of the interior $\mu$-metal shield, placed along the vacuum system of the interrogation cavity (brown shield of (a)).}
\label{fig:migaAIoverview}
\end{figure}

Each atom interferometer of the MIGA antenna (see Fig. \ref{fig:migaAIoverview}) uses clouds of $N\simeq 10^6$ atoms at a temperature of a few $\mu$K with a repetition rate of 1 Hz. Such sources are based on a 2D-MOT, which loads a 3D-MOT cooled to sub-Doppler temperature and launched vertically in a moving molasses, with a controlled velocity of $\approx$4 m/s. Before entering the interferometric region, the atomic clouds are prepared on the internal $m_F=0$ state, sensitive to magnetic fields only at the second order, and with a horizontal velocity that fulfills the Bragg condition for diffraction in the interferometric cavities. Two sets of horizontal Raman beams, tuned on the transition between the magnetically insensitive atomic sub-levels, are used to this scope, together with resonant optical beams to remove the atomic population on unwanted levels.

At the apex of their parabolic trajectory, the clouds experience a series of in-cavity $\pi/2$, $\pi$, $\pi/2$ pulses before returning in the detection region where the transition probability is measured. The Bragg interferometer uses the same internal state for both atomic paths, and the output ports are separated only for their horizontal velocity. A Raman pulse is then applied to reflect the horizontal velocity of one of the two output ports, and obtain internal state labeling for the two atomic populations. The counting on the two output ports uses fluorescence detection with atom shot noise sensitivity \cite{Santarelli1999,Rocco2014}. The normalized population ratio gives the transition probability of the interferometer, and then the atom phase shift.

The preparation phases for the atomic ensembles - cooling, trapping, launch, optical pumping and velocity selection - and the successive fluorescence detection make use of a commercial laser system based on telecom technology and frequency doubling to obtain the required light beams at 780 nm \cite{MUQUANS}.

\subsection{Cavity enhanced interrogation}\label{cavityInterro}

\begin{figure}[!h]
\centering
\includegraphics[width=0.9\textwidth]{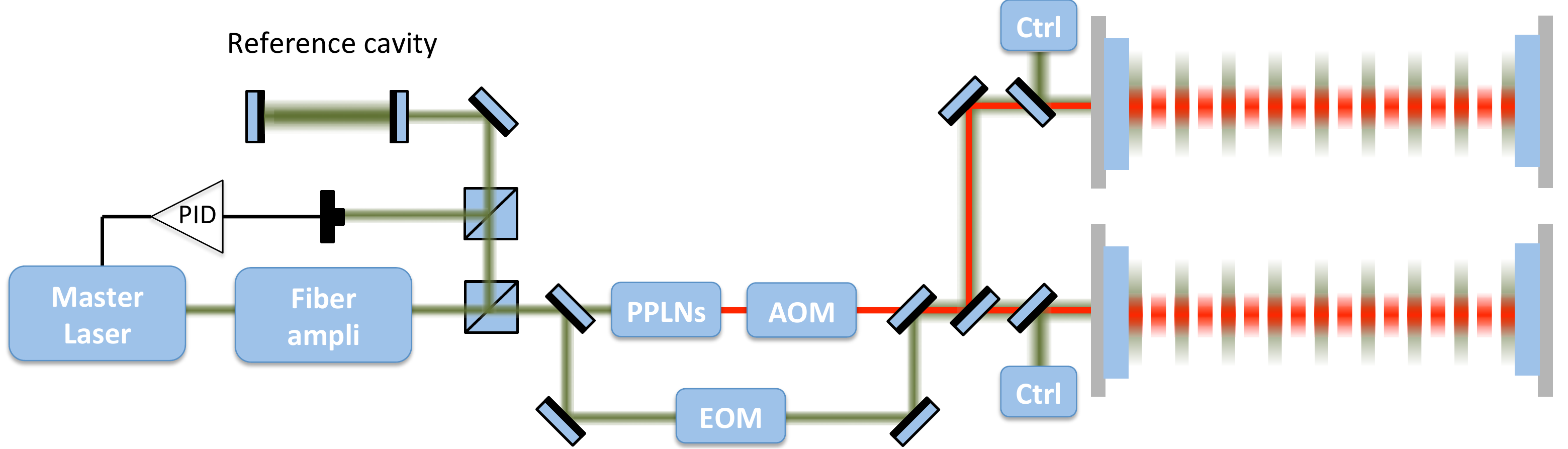}
\caption{Scheme of the laser interrogation system, from \cite{Canuel2016}. A master laser at 1560 nm (represented in green) is amplified and locked to a reference cavity, before being frequency doubled to obtain the interrogation radiation at 780 nm (represented in red). The master laser is frequency locked to one of the two interrogation cavities, whereas the component at 780 nm is pulsed by means of an AOM.}
\label{fig:laser_sys}
\end{figure}

Two horizontal cavities of 200 m are used to coherently interrogate the three atomic ensembles simultaneously launched along the vertical direction. A $\pi/2-\pi-\pi/2$ Bragg pulse sequence is applied in a symmetric fashion with respect to the apogee of the atomic parabolas: the matter wave splitting and recombination are implemented through the lower cavity via $\pi/2$ pulses, the redirection with the upper cavity via a $\pi$ pulse. The laser radiation to interrogate the atoms is obtained by injecting the fundamental transverse mode TEM$_{00}$ of each cavity, and the Bragg pulses are shaped with acousto-optic modulators (AOMs). The resonance condition of each interrogation beam to the corresponding cavity is granted by generating the 780 nm light via frequency doubling of a master telecom laser at 1560 nm with a Periodically Poled Lithium Niobate (PPLN) crystal \cite{Chiow2012,San2012}; the telecom laser is amplified, pre-stabilized to $\delta\nu / \nu_0=10^{-15}$ level at 1 Hz using an ultra-stable reference cavity, and finally injected in the two interrogation cavities, as shown in Fig. \ref{fig:laser_sys}.
In order to keep the resonance condition, the length of each cavity is locked at low frequency -outside the detector bandwidth- on the interrogation laser at 1560 nm. The frequency of the pulse shaping AOM at 780 nm is chosen so as to match the cavity resonance condition for the diffracted beam; such frequency is fixed by the differential optical length of the cavity at 1560 nm and 780 nm, because of the effect of the two reflective coatings.

Due to the cavities' long storage time with respect to the duration of the Bragg pulses, the interrogation fields will suffer some degree of deformation of their temporal amplitude profiles \cite{Fang2018}. This deformation scales with the cavity finesse, and it can have an adverse impact on the interferometers by increasing the minimum interaction time of the velocity-selective atomic transitions as well as their power requirements \cite{DovaleAlvarez:2017}. On the other hand, the cavities offer spatial filtering of the interferometric beams due to their frequency-dependent resonance conditions. This filtering, which also scales with the cavity finesse, effectively reduces the sensitivity of the interferometers to laser wavefront distortions, which are a leading source of noise in current state-of-the-art detectors \cite{Kovachy2015b}. The finesse of the cavities must therefore be chosen as a trade-off between these two cavity-induced effects on the atom optics pulses. 

For the 200 m MIGA cavities a finesse of 100 is chosen to strike a balance between the two effects for Bragg orders $n \leq 10$ \cite{DovaleAlvarez:2017}. The resulting cavity bandwidth yields a minimum interaction time 1.5 times larger than in the absence of the cavity for $n=10$, but below $1/\omega_r$ for all $n$ (with $ \omega_r = \hbar k^2 /2M $ the 2-photon recoil frequency and $M$ the mass of the atom). In this configuration, most higher order spatial modes will be optically suppressed at the $10^{-2}$ level. A lower finesse would lead to less dilation of the interaction time and slightly improved power enhancement of the short beam splitter pulses, especially for the higher-order diffraction processes, but would incur a worse optical suppression of higher order modes, partially negating the benefit of the cavity-assisted atom optics.

The radii of curvature (ROC) of the cavity mirrors are chosen to yield a beam waist sufficient to interrogate the atomic clouds as they thermally expand during the measurement. Furthermore, the resulting cavity configuration provides suppression of mode degeneracy for Hermite-Gauss modes up to order 15, taking into account manufacturing tolerances of the ROC. For the chosen ROC of 555 m, the $g$-factor of the resonator is 0.64 and the beam radius is 7.28 mm at the waist and 8.04 mm at the mirrors, providing a good margin over the cloud radius at the last interferometric pulse, which is expected to be $\simeq$5 mm for a 1 $\mu$K $^{87}$Rb ensemble at the end of the 2T = 500 ms pulse sequence. The resulting cavity configuration is robust to ROC deviations, although a reliable alignment sensing and control system will have to be implemented.

The two cavities share a common payload on each side to hold the mirrors, placed at a vertical distance of $\simeq$30.6 cm to have an interrogation time of T=250 ms. The impact of ground seismic noise on the position of the cavity mirrors will be reduced by means of an anti-vibration system.
The interest of this system is two-fold: it limits the related phase noise contribution on each atom interferometer to a negligible level with respect to the atomic shot noise contribution and insures that the cavity remains close to resonance inside the detection bandwidth. We are working on two different approaches: a passive system of mechanical filters to suspend each payload, or an active stabilization of each mirror position using piezoelectric actuators. The main constraints on the anti-vibration system are set by the seismic noise level at the chosen location (see Par. \ref{Par:lsbb_lowNoise}), and the response function of the atom interferometers to mirror acceleration noise (see Par. \ref{par:cavity}). Furthermore, relative length fluctuations of the two cavities must be controlled, since they would determine a shift of the readout atomic phase.\\


The systematic shift on the interferometric measurement due to the Coriolis effect is differentially canceled in the gradiometric measurement, if the launch directions for the atomic ensembles do not change between the experimental runs. The effect of a change between two sensors in the relative launch direction of an angle $\Delta \theta$ in the horizontal direction perpendicular to $\textbf{k}$ will induce a Coriolis contribution to the differential phase \cite{Bertoldi2006}

\begin{equation}
\phi_C \; \Big|_{n=1} =-2 \Omega_E \sin \alpha_{lat} k T^2 v_l \Delta \theta \simeq 470 \cdot \Delta \theta 
\end{equation}

where $\Omega_E$=72.9 $\mu$rad/s is the Earth's rotation rate, $\alpha_{lat}$ the latitude equal to 43.9$^{\circ}$ at LSBB. For a systematic effect below the atomic shot noise, equal to 1 mrad in terms of the interferometric differential phase, a launch stability at 1 $\mu$rad level is required for each sensor.

\subsection{Environment and control system}

The operation of each atom interferometer and of the cavity enhanced optical link implementing the atomic interrogation sets stringent environmental requirements. The cold atomic ensembles must be manipulated in regions evacuated to high vacuum in order to limit collisions with the thermal particles in the residual atmosphere. Differential vacuum techniques are adopted to obtain the vacuum levels required by different experimental regions: the typical pressure in the 2D-MOT cell is $\simeq$10$^{-8}$ mbar, optimized to achieve a good atomic flux to load the 3D-MOT; the 3D-MOT, the detection, and the interferometer regions are operated at a pressure below 10$^{-9}$ mbar to have a long lifetime of the atomic samples and reduce systematic effects related to the presence of background pressure. The vacuum level along the optical link results less demanding than what required at the interferometric region, if the phase front distortions caused by the residual index of refraction are taken into account. Nevertheless, a similar requirement is finally obtained when one considers the difficulty to implement differential vacuum between the respective regions to limit optical diffraction of the Bragg beams into the cavity and the related decoherence effect. The interferometric interrogation time T=250 ms translates into a vertical distance of $\sim$30.5 cm between the optical axis of the two cavities; a single vacuum tube with a diameter $\Phi$=0.5 m is chosen to host them, since a configuration with two separated tubes with a reduced diameter and conductance would make challenging the evacuation of the system to the required level. The target vacuum pressure in the different experimental regions will be obtained using vibration free pumping systems, relying on a combination of ion pumps and non-evaporable getter pumps. 

The magnetic susceptibility of Rb \cite{Steck2001} requires magnetic shielding of a factor >10$^4$ for the regions where the atoms are prepared and coherently manipulated \cite{Dickerson2012,Schuldt2015,KubelkaLange2016}. As can be seen in Fig. \ref{fig:migaoverview}, 3 different shields will screen each atom interferometer: one is dedicated to the atomic source and detection region, and two to the interferometric region.


The experimental sequence, timing and data acquisition are computer controlled. The synchronization of the three atomic sensors is granted by the use of common interrogation beams, which is the same element on which relies the high rejection ratio of common mode noise in the gradiometric configuration. On the other hand, the optimization of the three atomic signals requires a good synchronization of the preparation and launching sequence for each of the three atomic cloud, which will be obtained by referencing and precisely phasing the operation of each atomic head to a common time signal.

\section{\label{Par:LSBB}MIGA installation site: the LSBB underground science platform}

\begin{figure}[h!]
	\centering
	\includegraphics[width=0.74\textwidth]{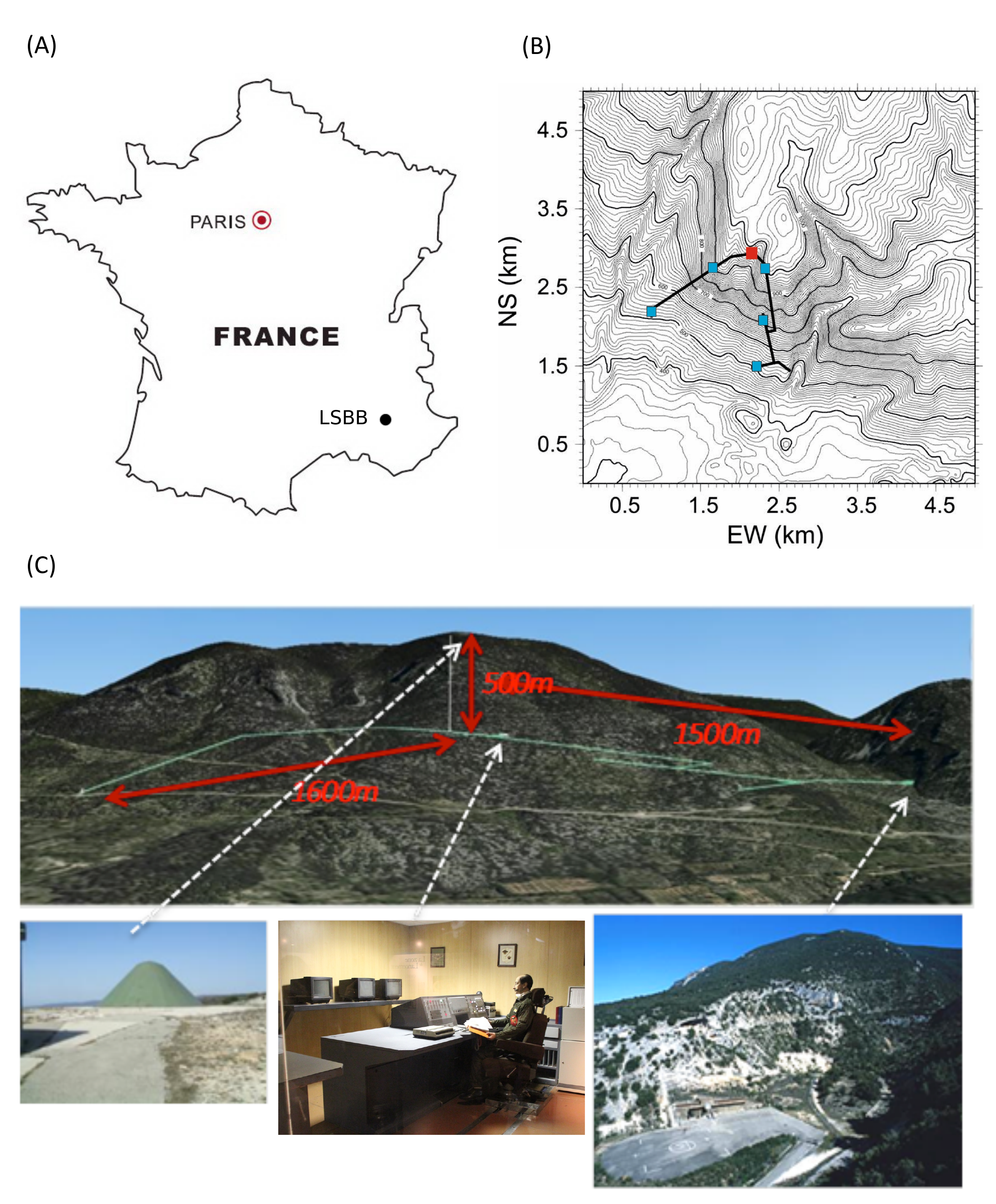} 
	\caption{(A): The LSBB is placed in south-east France. (B): Topographic map with the galleries (thick black lines) and permanent broadband seismic stations installed at the surface (red rectangle) and in the galleries (blue rectangles)(C): On the top image the 4-km LSBB galleries (cyan lines) and the length of its different branches (red). On the bottom images, from left to right: view of the top of the mountain with the shielded antenna dome intended to receive the launching order for the nuclear rockets; the launching control room in a EM shielded volume; the access to the galleries.}
	\label{fig:lsbb}
\end{figure}

MIGA will be installed at the LSBB, an underground low-noise laboratory located in Rustrel, near the city of Apt in Vaucluse, France (see Fig. \ref{fig:lsbb}). The LSBB is a European interdisciplinary laboratory for science and technology created in 1997 from the decommissioning of a launching control system of the French strategic nuclear defense operative during the Cold War. The LSBB is now an underground scientific platform \cite{Bettini2012} characterized by an ultra-low noise environment, both seismic and electromagnetic, as a result of the distance of the site from heavy industrial and human activities. The LSBB fosters trans-disciplinary interactions and interdisciplinary approaches, pursuing both fundamental and applied research. The result is a broad scientific and industrial expertise \cite{Gaffet2009}, besides the openness to European and international research programs and groups. The LSBB is an ideal facility for site studies of next generation GW detectors and more generally for the improvement of low-frequency sensitivity of existing and future GW antennas. In such environment, MIGA aims at studying Newtonian Noise and testing advanced detector geometries for its cancellation \cite{Chaibi2016}.

The LSBB is embedded in the Fontaine-de-Vaucluse watershed, which is one of the
world's largest karst aquifers \cite{Ford2007}, and covers 54 ha in surface area. The underground facility includes 4 km-long horizontal drifts at a depth ranging from 0 m to 518 m, with north to south and north-east to south-west as its main orientations (Fig. \ref{fig:lsbb}). The galleries give access to underground wells, vaults and voids, where a passive temperature stability better than 0.1 $^o$C results from the thermal blanketing supplied by the upper carbonate rock layer.  Air pressure and circulation are controlled in the galleries by airlocks. At the deepest point, 518 m under the surface, is located the 1,250 m$^3$ vault previously equipped as nuclear launch control room, hence designed to remain operational even in the case of a nearby nuclear blast. This place is nowadays used for the most demanding experiments in terms of seismic and electromagnetic residual noise. The whole system of underground drifts and surface areas is connected to energy, telephony, optical fiber Internet, and GPS synchronization.
The LSBB's environment provides benefits for developing ultra-sensitive instrumentation, calibration and comparison of highly sensitive sensors, optimization of experimental protocols and the operation of industrial prototypes. The LSBB takes advantages of its remote location and its robust design and provides access to the karstified carbonate reservoir through galleries that may be assimilated to kilometric horizontal boreholes. The site provides an original access to the karst medium, complementary to usually investigated caves and springs \cite{Garry2008}. Therefore, it is of broad interest for hydrogeologists and geophysicists who study karst systems \cite{networkHydro,Carrire2016,Lesparre2016}. Such a configuration helps also to develop research programs on poro-elastic dynamics of fractured media, induced seismicity and processes for internal damaging of rock massifs, and the handling of fluids and gas in reservoir \cite{Wang2010}. A broad range high sensitivity instruments (e.g. optical strainmeters, a hydrostatic long baseline tiltmeter, seismometers, superconducting magnetometers and gravimeters, muon cameras for rock densimetry, cold atom gravimeters) is studied and developed at LSBB; their metrological performances are characterized and cross-compared. The interdisciplinary ability of the LSBB allows to take into account the effects induced by the disturbances of the physical environment of the sensor on the measure it produces, like the rock mass tilts \cite{Lesparre2016} or influence of gravity gradient variations on deformation measurements \cite{Deville2012,Fores2016}

\subsection{\label{Par:lsbb_lowNoise}LSBB low noise properties}

The underground environment of LSBB is characterized by exceptional low-noise properties, which is the basic condition to realize high sensitivity experiments in diverse fields, like for example the first observation of simultaneously recorded seismic and seismo-magnetic signals caused by an earthquake \cite{Gaffet2003}, the acquisition of extremely weak biological signals \cite{Zandi2011}, and the operation of a portable atomic gravimeter at high sensitivity without an anti-vibration platform \cite{Farah2014}.

\begin{figure}[h!]
	\centering
	\includegraphics[width=1.0\textwidth]{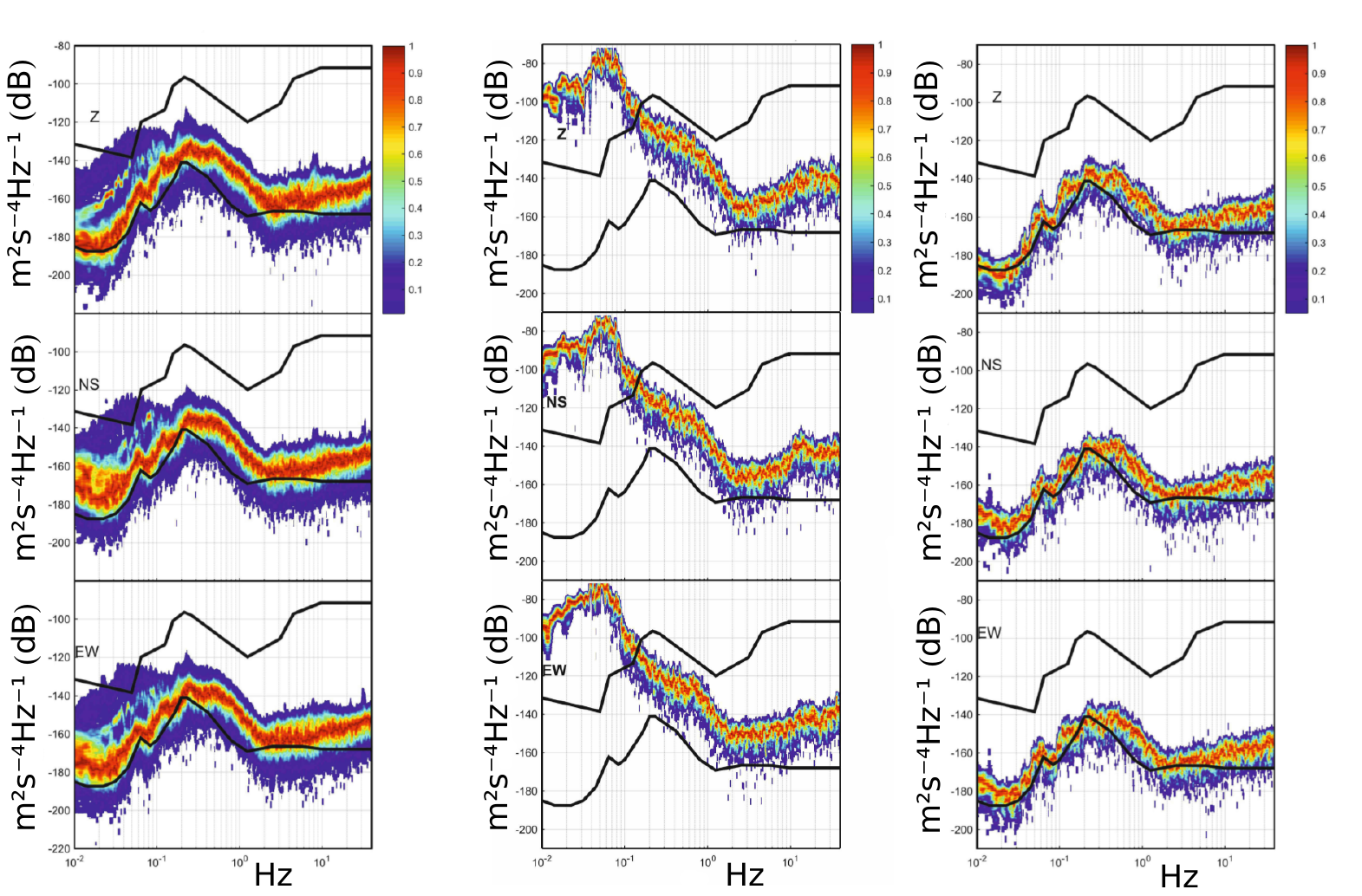} 
	\caption{Probability density function of seismic ambient noise (color scale) of RUSF.01 broadband station for year 2011, including quiet days and days with earthquakes (left); 6 hours of ground motion which includes the Tohoku-Oki earthquake of March 11, 2011 (center); a quiet day (24 hours) of year 2011 without seismic events (right). The seismic noise PSD for three components (top, Z; middle, NS; bottom, EW) is compared to the Peterson's high and low noise models (black lines).}
	\label{fig:pdf_lsbb}
\end{figure}
A network of 6 broadband seismometers is deployed at LSBB to monitor the seismic noise variations at the site \cite{Gaffet2009}. Sample acceleration noise signals recorded at the station RUSF.01, located underground in a very quiet place, are presented in Fig. \ref{fig:pdf_lsbb}: their probability density function (PDF) is compared to the Peterson's models \cite{Peterson1993}, commonly used as a reference for the definition of the quality of a seismic recording site. The seismic data for three orthogonal components are reported for three different measurement intervals: the whole 2011 year, which includes the Mw 9.1 Tohoku-Oki March 11, 2011 mega-thrust Japanese earthquake; a 6 hour interval during which the same event took place; a 24 hours interval during 2011 without seismic events. The graphs shows that the site has remarkable seismic properties, and, except during important transient signals, the highest probability of noise occurrence over a long interval is close to the new low-noise Peterson's model for all the three components and for the whole considered frequency band (Fig. \ref{fig:pdf_lsbb}-left). At frequencies below 0.2 Hz, the one-year PDF is broader and spreads between the high and low noise models of Peterson, since it includes the worldwide seismic activity and hence surface waves with common frequencies below 1/20 Hz and high amplitudes; among all events the strongest one has been the Tohoku-Oki March 11 earthquake, for which high seismic energy was measured at low frequency (Fig. \ref{fig:pdf_lsbb}-center). The noise PDF over a quiet day gives a sharp high probability that at low frequencies borders (horizontal components) or even drops below (vertical component) the low noise model of Peterson (Fig. \ref{fig:pdf_lsbb}-right). 

\begin{figure}[h!]
	\centering
	\includegraphics[width=0.35\textwidth]{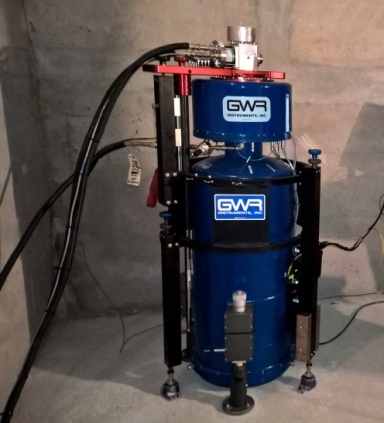} 
	\includegraphics[width=0.61\textwidth]{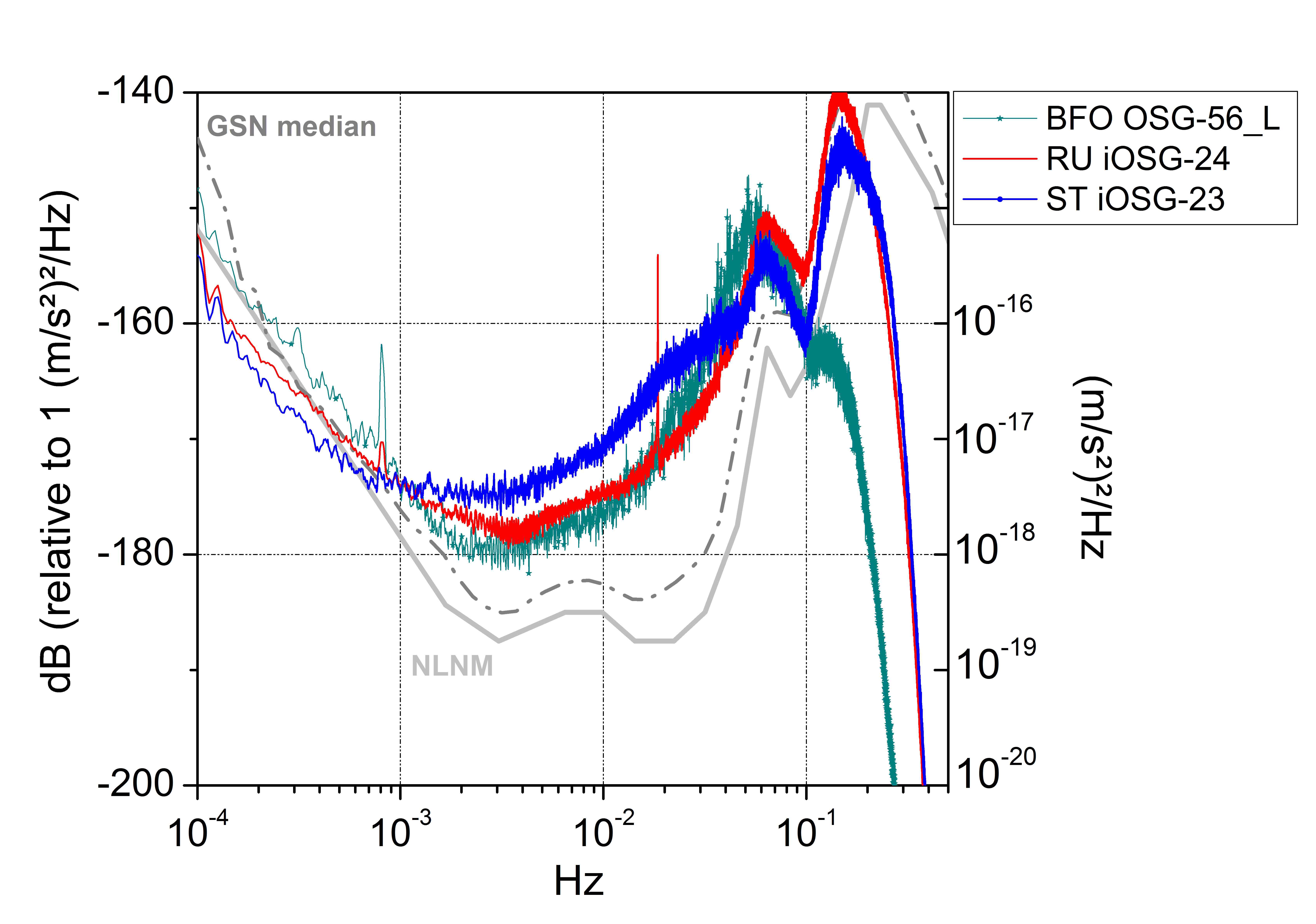} 
	\caption{(left) The superconducting gravimeter installed at LSBB in the frame of MIGA experiment. (right) Comparison of the noise power spectral density for superconducting gravimeters at different world sites: LSBB at Rustrel - France (``RU'', red); at Strasbourg - France (``ST'', blue); Black Forest Observatory near Schiltach - Germany (``BFO'', green). The signals are obtained from daily power spectral densities on raw data sampled at 1 second. The sharp drop at high frequency (>0.1 Hz) is due to the anti-aliasing filters present in the superconducting gravimeters. The Peterson's low noise model (NLMN in solid gray) and the seismological GSN median noise model of Berger \cite{Berger2004} (GSN median in dashed-dotted gray) are plotted for reference.}
\label{fig:graviComparison}
\end{figure}

The low-noise seismic properties of LSBB have been confirmed also with gravity measurements realized with a high sensitivity superconducting gravimeter (model iOSG from GWR Instruments Inc., Fig. \ref{fig:graviComparison}-left), installed at the underground site in 2015 as a complementary instrument to the MIGA experiment. Fig. \ref{fig:graviComparison}-right shows a recent measurement of the noise level at LSBB, compared with the curves obtained in Strasbourg (France) and at the Black Forest Observatory (Germany). The instrument installed at LSBB has a noise performance among the best in a worldwide network of superconducting gravimeters \cite{Rosat2016}, with an amplitude spectral density of 1.8 nm/s$^2$/Hz$^{-1/2}$ at 1 mHz. Notably, superconducting gravimeters have a characteristic noise lower than the Earth’s background noise at frequencies below 1 mHz \cite{Rosat2011}, and permit thus to study low-frequency seismic and sub-seismic modes. 

The specific location of LSBB beneath a massif determines a unique sheltering effect as concerning electromagnetic noise and its effect on the measurement of the atom interferometric phase \cite{Canuel2007}. The laboratory lies in a very quiet environment, the Regional Natural Park of Luberon: minimal human activities within two kilometers reduce the magnetic interference from railways and high-voltage power lines. Despite being in a moderately seismic area, the seismic noise spectra recorded at the site are close to the worldwide minima, hence there is no significant movement of magnetic mass, like the content of the nearby water reservoir, which could perturb the magnetic background noise. 

An exceptionally low noise electromagnetic environment is provided by the former control room of the military facility: such infrastructure was designed to withstand the effects of a nearby nuclear blast, including the electromagnetic pulse (EMP) in its deepest zone. The room consists of an unconventional Faraday cage, where the electromagnetic shielding effect is provided both by karstic 500 m thick rocks loaded with water, which gives a high-frequency cut-off around 200 Hz, and by the 2 m thick reinforced concrete capsule with a 1 cm thick steel inner coating, which reduces the cut-off to 10 Hz. The absence of $\mu$-metal makes it a perfect low frequency pass-band filter, which is different from a zero Gauss chamber. The performance of the shielding was measured with a 3-axis Superconductive Quantum Interference Device (SQUID) magnetometer. The residual noise level is about 2 fT/$\sqrt{\mathrm{Hz}}$ above 40 Hz \cite{Waysand2000}, and the attenuation in the DC domain leads to a residual field inside the Capsule of about 6 $\mu$T, compared to the 46 $\mu$T expected at the LSBB latitude. Inside the vault is placed a cabin measuring 20$\times$6 m$^2$ and weighting 30 t, hangs from the ceiling and rests on eight shock absorbers ensuring the decoupling from the ground movement. This location is used to perform experiments that require the most demanding quality for the seismic and electromagnetic environment.

\begin{figure}[h!]
	\centering
	\includegraphics[width=1.0\textwidth]{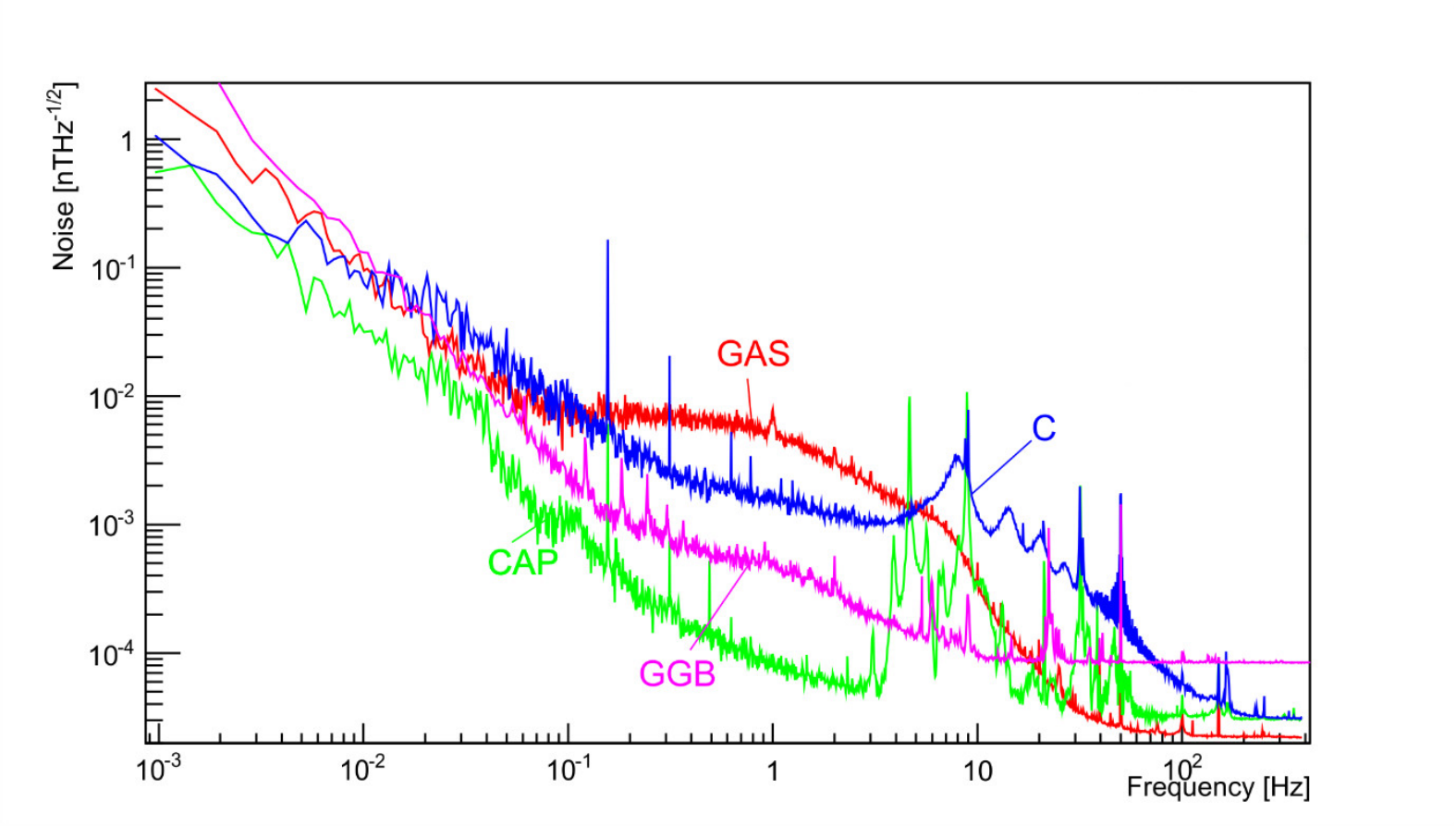}
	\caption{From \cite{Henry2016}. Frequency spectra of magnetic fluctuations measured at four different locations in the LSBB complex, indicated on the map in Fig. \ref{fig:lsbb_miga}: the Capsule (CAP), the Anti-blast Gallery (GAS), the Secondary Gallery (GGB), and the Safety Gallery (C).}
	\label{fig:noiseEM}
\end{figure}

Fig. \ref{fig:noiseEM} shows frequency spectra recorded in the Capsule using a portable SQUID magnetometer, and at three other points inside the tunnel complex. In the range 0.01-10 Hz of interest to MIGA, all show a similar level till 60 mHz, except at the Capsule where the steel walls attenuate the geomagnetic fluctuations. Above 0.1 Hz, the magnitude of the fluctuations varies with the screening of the surrounding rock and any interference from nearby equipment. The spectra recorded at point C show broad peaks at the Schumann resonance frequencies. Other peaks can be attributed to mechanical vibrations or electromagnetic interference (50 Hz power lines and its harmonics are clear). At higher frequencies, the instrument noise starts to dominate. The signals are screened by the telluric currents induced in surround rock by the changing external field. This creates a significant gradient in these magnetic signals across the laboratory. A search for magnetic field changes correlated with large changes in groundwater flow has ruled out signals above 0.2 nT \cite{Henry2014}.

\subsection{MIGA infrastructure at LSBB}

\begin{figure}[h!]
	\centering
	\includegraphics[width=0.71\textwidth]{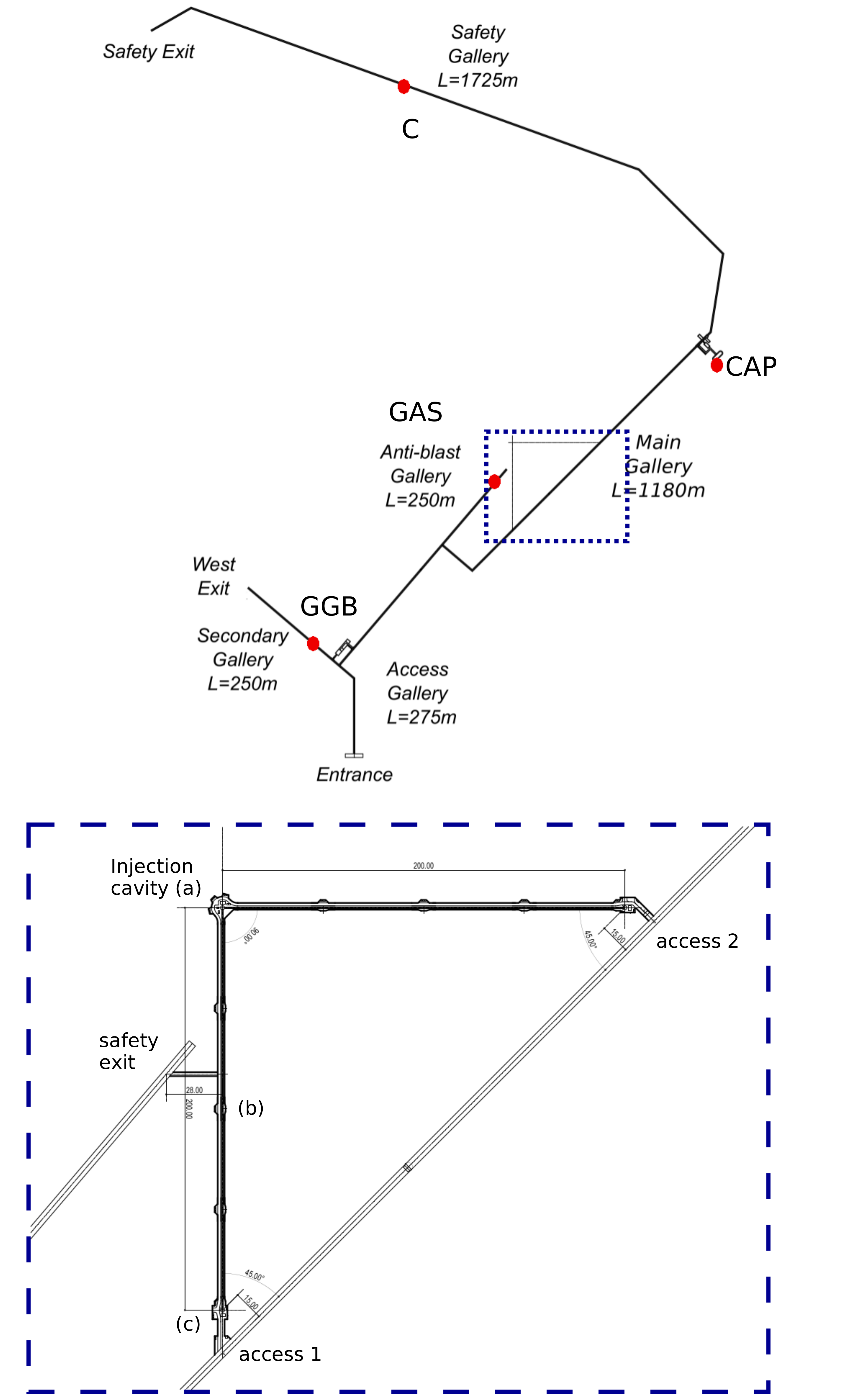} 
	\caption{(Above) Map of the underground galleries at LSB, with the locations adopted for the characterization of magnetic fluctuations at the site indicated by red points. The place where MIGA will be installed is highlighted with a dashed blue rectangle. (Below) Zoom of the MIGA infrastructure: the two orthogonal boreholes will use the main gallery at their far ends for access, whereas the anti-blast gallery will be used as a safety exit. The three AIs will be located in the room use to inject the cavities (a), at the other end of one gallery (c) and at its mid position (b).}
	\label{fig:lsbb_miga}
\end{figure}

The initial approach of installing MIGA in the ``Main Gallery'' at LSBB (see Fig. \ref{fig:lsbb_miga}) has been discarded, in relation to the 5\% slope of the tunnel, which will introduce the projection of the gravitational acceleration into the measurement, and also to the requirement to have a quiet dedicated site and not a shared environment used to reach other experiments. 
Two horizontal galleries will be instead bored to host the MIGA instrument, as shown in Fig. \ref{fig:lsbb_miga}. The choice to have two orthogonal tunnels is related to the possibility to mitigate the impact of the interrogation laser technical noise on strain measurements by adopting a Michelson-Morley configuration, in the same way as LIGO and VIRGO do. Nevertheless, mainly for budgetary reasons, only one tunnel will be initially equipped with an atom--laser antenna. The second tunnel will be equipped later to realize a 2D gravitational antenna. The orientation and position of the tunnels is chosen so as to exploit the existing galleries for the required access points at their extremities for safety reasons. This configuration forbids the use of drilling machines for the excavation, which will be realized using explosive charges. The explosions will be exploited by geoscientists at LSBB to study the response of the karstic mountain to man-made seismic activity. Boring the two 200 m long, 3.2 m wide tunnels will require a 12-18 months period of time, depending on the quality of the rocks, and an estimated cost of about 4 M\euro.

\begin{figure}[h!]
	\centering
	\includegraphics[width=\textwidth]{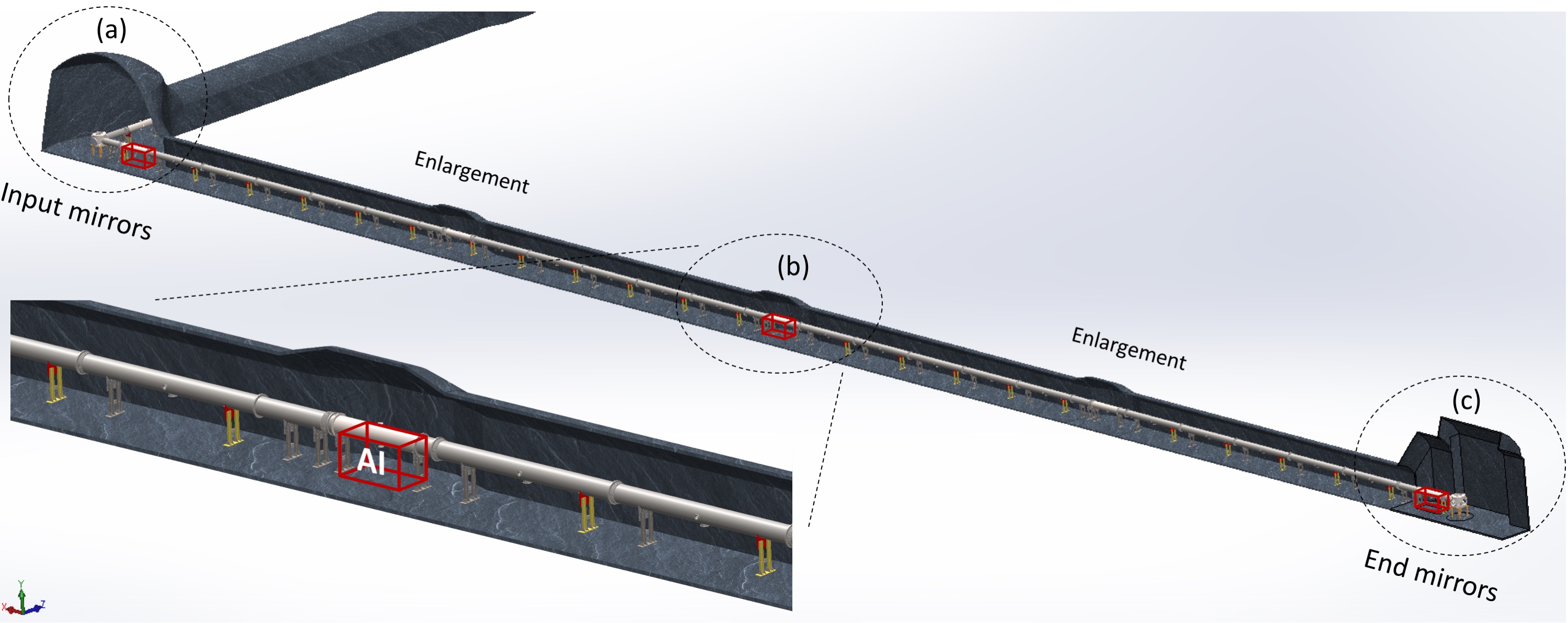} 
	\caption{Design of the galleries dedicated to the MIGA experiment at LSBB. The three atom interferometers of the antenna will be located at (a), (b) and (c). The optical setups for cavity injection will be hosted in room (a). The two MIGA galleries will be equipped with enlargements regularly spaced in order to add other AIs in the future.}
	\label{fig:lsbb_infra}
\end{figure}

A preliminary design of the infrastructure is presented in Fig. \ref{fig:lsbb_infra}; the size of the two tunnels and of the auxiliary rooms required for the cavity injection and for the atom interferometers is defined in terms of the instrument encumbrance, safety and environmental requirements. The tunnels will have to host the vacuum system used by the optical link, the AIs, the cavity injection optical setups and various electronic and data acquisition equipments. Apart from the large cavities at each end of the galleries, widenings will be present every 50 m along the tunnel to have the possibility to change the distance of the three initial AIs, and that of increasing the number of AIs so as to improve the spatial resolution of the experiment.

\subsection{Geophysics at LSBB with MIGA}

MIGA will extend the concept of correlated interferometry from the laboratory scale to that of a geological site, the LSBB at Rustrel, using an underground array of atom sensors distributed along a 200 m horizontal arm. Several techniques based on correlated atom interferometry will be implemented to characterize the gravitational field of the site, such as the simultaneous measurement of gravity acceleration and gradient \cite{Sorrentino2012} and the measurement of gravitational curvature \cite{Rosi2015,Asenbaum2017}. It will be thus possible to investigate several geological phenomena, like the non-invasive detection of underground density anomalies \cite{Metje2011}, the gravity perturbations due to local density changes caused by fault evolution as proposed in \cite{Harms2013}, and the characterization of gravity-gradient noise, also called Newtonian Noise (NN) \cite{Chaibi2016,Canuel2016}.

The MIGA differential signal is sensitive to gravity gradient $x_I(X_i)-x_I(X_j)$, as can be seen in Eq. \ref{AIresponse3}. The fluctuation of this term determines the NN, a tidal effect that affects any couple of test masses, both macroscopic and microscopic, and is considered to be a fundamental limit for any ground based GW detector. Nevertheless, GW and NN have different length scales over the detector's dimensions: GW signals have extremely long characteristic lengths and are seen as pure gradients at the distances of interest, whereas NN has shorter characteristic lengths going from the meter to a few kilometers \cite{Saulson}. Effects due to NN may become discernible using a network of correlated sensors distributed along the antenna's direction \cite{Chaibi2016}, which could then pave the way to novel rejection methods for this kind of disturbance and ultimately to the realization of sub-Hz ground based GW detectors.

MIGA will provide absolute gravity and gravity-gradient measurements, which will be used to obtain density maps for the surrounding volume via inversion algorithms. The limited resolution of the measurement, due to the reduced number of atomic sensors along the cavity arm, their fixed position and directionality, will be mitigated by correlating the AI measurements with other kinds of gravity measurements provided by the instrument network deployed at LSBB. This approach will also implement a long baseline hybrid gravity sensor, broadening the concept introduced in \cite{Lautier2014}. The device will be calibrated on the signal produced by heavy source masses \cite{Gray1995,Lamporesi2007} placed at a variable distance up to 150 m, exploiting the nearby tunnels already present at the site. The instrument will provide a gravity gradient sensitivity of the order of 10$^{-3}$ E{\"o}tv{\"o}s at 1 s (1 E = 10$^{-9}$ s$^{-2}$), an order of magnitude below the reported sensitivity of superconducting devices over much smaller probe distances \cite{VolMoody2002}. The absolute readout of the AI-based device would finally allow to periodically calibrate the other geophysical instruments operated at LSBB.

In the following phase, the underground structure of the surrounding karstic environment will be studied, to detect subsurface cavities \cite{Butler1984} or monitor groundwater dynamics. Indeed, better characterization of complex underground reservoirs is expected from recent and future developments of geophysical methods \cite{Berkowitz2002}. Their application to karst is probably the most challenging \cite{Chalikakis2011} because karst heterogeneity is multi-scale, self and hierarchical organized and induces the most complex underground fluid dynamics. Peculiarly, the lack of non-invasive methods producing multi-scale 4D imaging of underground fluids remains a bottleneck for understanding and modeling this dynamics. One of the important questions is to have enough resolution and depth of investigation all at once to detail all the features controlling the groundwater circulation and storage from matrix porosity or micro-fracturing to major faults and karst conduits. Currently, only integrating methods directly or indirectly related to water content such as seismic, Electrical Resistivity Tomography (ERT), Magnetic Resonance Sounding (MRS) or gravimetry allow estimating the variation of water mass. Since no method presents at the same time the required resolution, depth of investigation and fluids sensitivity, coupling and comparisons appear to be a promising way for imaging underground structure and fluids dynamics (e.g. \cite{Carrire2013,Carrire2016}). Thus the place of gravimetry surveys is increasing in hydrological studies \cite{Hector2014,Hasan2008,Pool1995,Deville2012}. Combining conventional instruments and methods from hydrogeology with cold atom gravitation sensor measurements will allow better understanding and modeling of karst aquifers \cite{Geiger2015}, for which 4D data are currently lacking to constrain spatially distributed models \cite{Ghasemizadeh2012}.

With MIGA we plan also to apply aperture synthesis imaging techniques to geophysics, adopting well developed techniques from radio-astronomy \cite{Thompson2007} and optics \cite{Greenaway1991}. To this aim, the distance between the atomic sensors installed on the optical link will be changed during successive measurement runs, so as to probe different Fourier components of the gravity field.

\color{black}

\section{Conclusions and outlook}

We presented the underground MIGA atom-laser antenna that is being built at LSBB in Rustrel to measure space-time strain in the infrasound with an expected peak sensitivity of 2$\cdot 10^{-13}/\sqrt{\mathrm{Hz}}$ at 2 Hz. The experiment is presented in what will be its environment, characterized by an exceptionally low seismic and electromagnetic noise, and by the presence of a network of geophysical instruments monitoring the site. The setup will represent a demonstrator for gravitational wave detection using atom interferometry, in a frequency band not explored by classical ground and space-based observatories

and will also permit to measure geophysical phenomena with unprecedented sensitivity. The underground location of the setup as well as its size pose several technological and scientific challenges, which are addressed in the article together with the devised solutions. However, this location offers a great opportunity to confront MIGA measurements with other gravity monitoring sensors and to explore innovative interdisciplinary applications of atom interferometry, especially in geoscience. After the first phase, described in this article, the instrument and specifically its strain sensitivity will be upgraded by adopting several advanced techniques, like large momentum beamsplitters \cite{Chiow2011,McDonald2013,Estey2015}, non-classical input states \cite{Hosten2016,Cox2016}, the asynchronous interrogation of many ensembles \cite{Dutta2016}, and the increase of interrogation time via optically guided interferometry \cite{McDonald2013a} or intermediate coherence preserving interrogations and feedback \cite{Kohlhaas2015}.

\renewcommand{\thesection}{\Alph{section}}
\setcounter{section}{0}

\section{Calculation of the intracavity field frequency modulation under the influence of a monochromatic GW.}\label{AnnexA}
\begin{figure}[!h]
	\centering
\includegraphics[width=7cm]{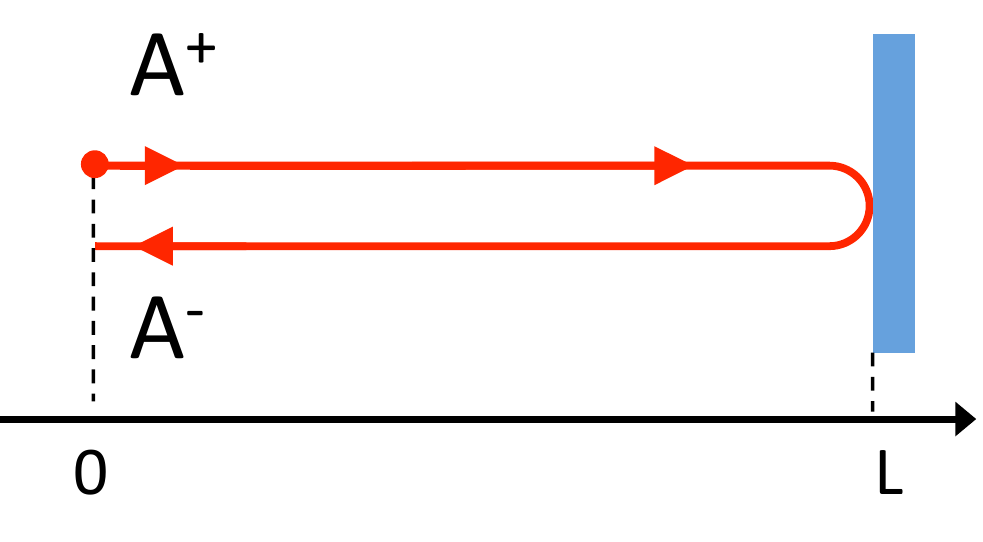}
\caption{A laser field $A^+$ is emitted at position 0 and reflected at distance L. $A^-$ is the returning field.}
\label{AllerRetour}
\end{figure}

\begin{figure}[!h]
	\centering
\includegraphics[width=7cm]{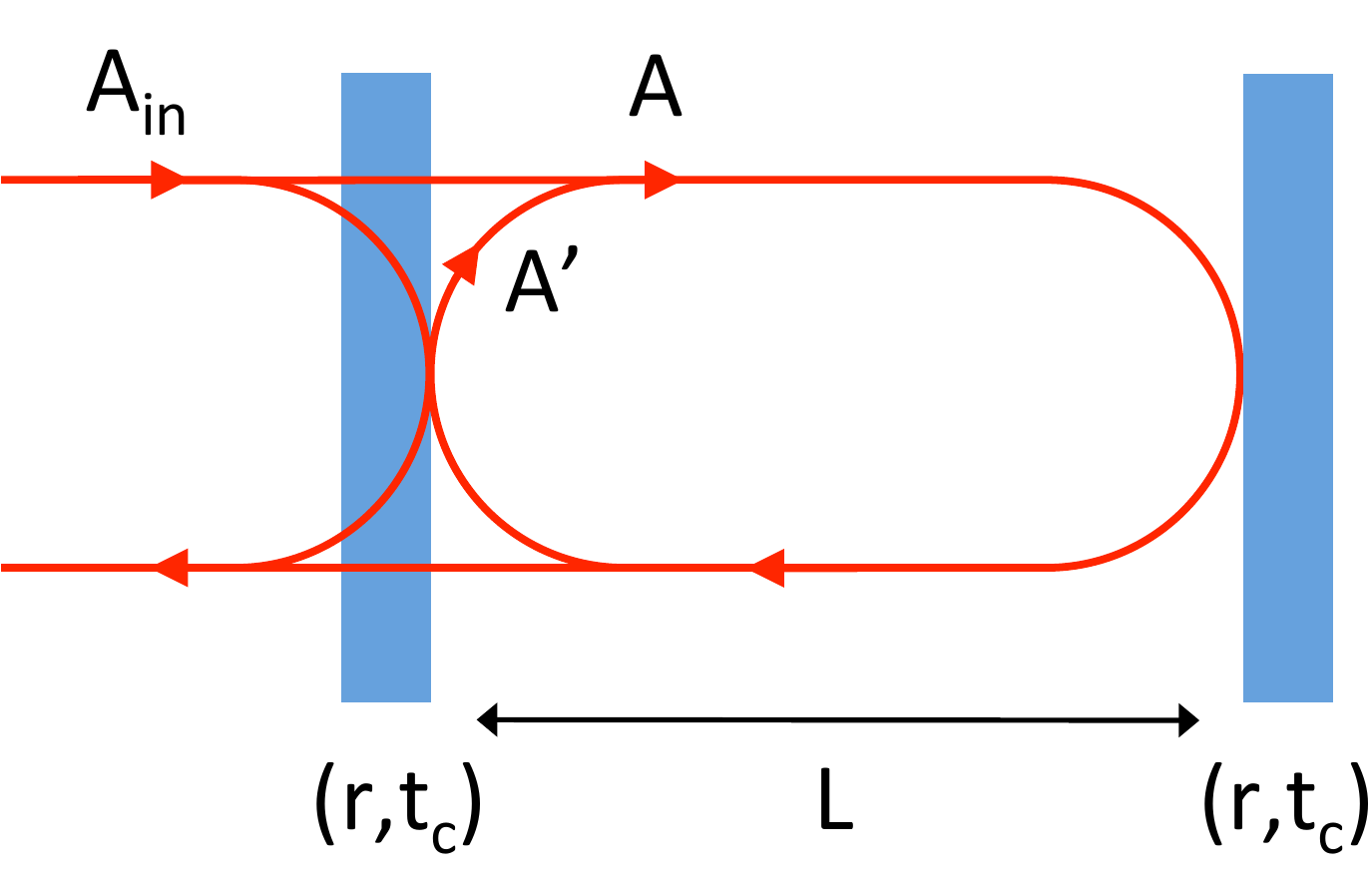}
\caption{A laser field $\mathbf{A_{in}}$ is incident on a cavity of length L formed with two mirrors with coefficients of reflection and transmission $(r,t_c)$. $\mathbf{A}$ is the circulating field and $\mathbf{A}'$ its copy after one round trip.}
\label{caviteparam}
\end{figure}

We first consider the effect of a monochromatic GW $h(t)=h \cos \Omega_{gw} t$ on a single round-trip of distance $2L$. We consider that a laser field $A^+(t)=A_0e^{-i\omega t}$ is emitted at position 0 (see Fig. \ref{AllerRetour}) and reflected at distance L. 
When it comes back to point with position 0, the field $A^-(t)$ can be expressed: 
\begin{equation}
A^-(t)= A^+(t_r)
\end{equation}
where $t_r$ is the retarded time calculated in Sec. \ref{par:cavity} (see Eq.\ref{eq:RetartedTime}) which can be expressed to first order in $\Omega_{gw}L/c$: 
\begin{equation}
t_r=t-\frac{2L}{c}+h\frac{L}{c} \cos (\Omega_{gw}(t-L/c)).
\label{RetardedTimeGW}
\end{equation}
 $A^-(t)$ results to be:
\begin{equation}
A^-(t)= A_0e^{-i\omega(t-\frac{2L}{c})}e^{-ih\frac{\omega L}{c}\cos (\Omega_{gw}(t-L/c))}.
\end{equation}
Expanding the second exponential term in h,  one obtains to first order:
\begin{equation}
A^-(t)= A_0e^{-i\omega(t-\frac{2L}{c})}\left(1-ih\frac{\omega L}{c}\cos \left(\Omega_{gw}(t-L/c)\right)\right)
\end{equation}
and then:
\begin{equation}
A^-(t)= A_0e^{-i\omega(t-\frac{2L}{c})}-\frac{1}{2}ihA_0\frac{\omega L}{c}e^{2i\frac{\omega L}{c}}e^{i\frac{\Omega_{gw} L}{c}}e^{-i(\omega+ \Omega_{gw})t}-\frac{1}{2}ihA_0\frac{\omega L}{c}e^{2i\frac{\omega L}{c}}e^{-i\frac{\Omega_{gw} L}{c}}e^{-i(\omega- \Omega_{gw})t}.
\end{equation}
After one round trip, the effect of the GW is to introduce two sidebands at frequencies $(\omega \pm \Omega_{gw})/2\pi$ of amplitude proportional to h. 
We calculate the effect of a GW on the circulating field of a cavity following closely \cite{VPB2006} in Sec. 1.3. Going back and forth many times inside the cavity, the intracavity field will remain to first order in h of the form:
\begin{equation}
A(t)=\left(A_0+\frac{1}{2}hA_1e^{-i\Omega_{gw}t}+\frac{1}{2}hA_2e^{i\Omega_{gw}t}\right)e^{-i\omega t}
\label{ExpressionAt}
\end{equation}
which enables to defines a "generalized amplitude" for the cavity field as a rank 3 vector $\mathbf{A}$:
\begin{equation}
\mathbf{A}=(A_0;A_1;A_2)
\label{}.
\end{equation}
We consider in the following a field $\mathbf{A_{in}}=(A_0;0;0)$ incident on a cavity of length L formed with two mirrors with coefficients of reflection and transmission $(r,t_c)$ (see Fig. \ref{caviteparam}). 
The amplitude of the cavity field after one round trip $\mathbf{A'}$ can then be expressed \footnote{E$_{in}$,E$_{ref}$ and E$_{trans}$ are respectively the incident, reflected and transmitted amplitudes on a cavity mirror; we obtain $E_{ref}=e^{i\pi/2}rE_{in}$ and $E_{trans}=t E_{in}$
} as a matrix product: 
\begin{equation}
\mathbf{A'}=-r^2\textbf{X}\mathbf{A}
\label{}
\end{equation}
where $\textbf{X}$ is the linear round trip operator:
\begin{equation}
\textbf{X}= e^{2iu}\left(
\begin{array}{ccc}
  1& 0  & 0  \\
  -i u \mathrm{sinc}(v) e^{iv}&e^{2iv} & 0  \\
  -i u \mathrm{sinc}(v) e^{-iv}&  0 &   e^{-2iv}
\end{array}
\right) 
\label{}
\end{equation}
where $u=\frac{\omega L}{c}$ and $v=\frac{\Omega_{gw} L}{c}$.
Considering only the first order in $v$ of $\mathrm{sinc}(v)$, one obtains:
\begin{equation}
\textbf{X}= e^{2iu}\left(
\begin{array}{ccc}
  1& 0  & 0  \\
  -i u e^{iv}&e^{2iv} & 0  \\
  -i u e^{-iv}&  0 &   e^{-2iv}
\end{array}
\right) 
\label{}.
\end{equation}
At steady state, the interference of the intracavity waves on the first mirror can be written (see Fig. \ref{caviteparam}):
\begin{equation}
\mathbf{A}=t_c\mathbf{A_{in}}-r^2\textbf{X}\mathbf{A}
\label{}
\end{equation}
which gives:
\begin{equation}
\mathbf{A}=[1+r^2\mathbf{X}]^{-1}t_c\mathbf{A_{in}}
\label{}.
\end{equation}
After Inversion of the $1+r^2\mathbf{X}$ Matrix, we calculate the generalized amplitude of the resonating field:
\begin{equation}
\mathbf{A}=A_0\left(\frac{t_c}{1+r^2 e^{2 i u}},\frac{i r^2 t_c u e^{i (2 u+v)}}{\left(1+r^2 e^{2 i u}\right) \left(1+r^2 e^{2 i
   (u+v)}\right)},\frac{i r^2 t_c u e^{i (2 u+v)}}{\left(1+r^2 e^{2 i u}\right) \left(r^2 e^{2 i u}+e^{2 i
   v}\right)}\right).
\end{equation}
The cavity being at resonance, we have $e^{2 i u}=-1$. From Eq. \ref{ExpressionAt}, $A(t)$ can then be written:
\begin{equation}
A(t)=A_0\left(\frac{t_c}{1-r^2}-\frac{i h r^2 t_c u e^{i t \Omega_{gw}+i v}}{2 \left(1-r^2\right) \left(-r^2+e^{2 i v}\right)}-\frac{i h r^2 t_c u e^{i v-i t
   \Omega_{gw}}}{2 \left(1-r^2\right) \left(1-r^2 e^{2 i v}\right)}\right)e^{i\omega t}.
\end{equation}
The resonating field can then be expressed in the form $A(t)=A_0e^{i(\omega t+\phi(t))}$ where:
\begin{equation}
\tan \phi(t)\simeq\phi(t)=\frac{h r^2 u \left(r^2\cos (v+t \Omega_{gw})-\cos (v-t \Omega_{gw})\right)}{r^4-2 r^2 \cos (2 v)+1}.
\end{equation}
To first order in $v$:
\begin{equation}
\phi(t)=h u\frac{r^2}{r^2-1}\left(\cos (t \Omega_{gw})-\frac{\left(r^2+1\right)}{r^2-1}v \sin (t \Omega_{gw})\right).
\end{equation}
Substituting $\frac{r^2+1}{r^2-1}\approx -\frac{2 F}{\pi }$ and $\frac{r^2}{r^2-1}\approx-\frac{F}{\pi }$ where $F$ is the Finesse of the cavity, we obtain:
\begin{equation}
\phi(t)=-\frac{FL}{\pi c}\omega h \left(\cos (t \Omega_{gw})+\frac{2F\Omega_{gw} L}{\pi c} \sin (t \Omega_{gw})\right).
\end{equation}
The frequency noise $\delta\nu_{gw}(t)$ induced by the GW on the resonating field can then be expressed as:
%
\begin{equation}
\delta\nu_{gw}(t)=\frac{1}{2 \pi}\frac{d \phi(t)}{dt}=\frac{F\Omega_{gw}L}{\pi c}\nu_0 h \left(\sin (t \Omega_{gw})-\frac{2F\Omega_{gw} L}{\pi c} \cos (t \Omega_{gw})\right)
\end{equation}
\begin{equation}
\delta\nu_{gw}(t)=\frac{1}{2}\frac{\Omega_{gw}}{\omega_p }\nu_0 h \left(\sin (t \Omega_{gw})-\frac{\Omega_{gw}}{\omega_p }\cos (t \Omega_{gw})\right)
\end{equation}
where $\omega_p/2\pi=c/(4LF)$ is the frequency pole of the cavity. For $\Omega_{gw}\ll\omega_p$ we approximate to first order:
\begin{equation}
\delta\nu_{gw}(t)=\frac{1}{2}\frac{\Omega_{gw}}{\omega_p }\nu_0 h \sin (t \Omega_{gw}).
\end{equation}

\section{Calculation of the intracavity field frequency modulation under the influence of a monochromatic cavity mirror vibration.}\label{AnnexB}
As in the previous Annex, we first consider the effect of a monochromatic mirror vibration  $\delta x(t)=\delta x\cos \Omega_{x} t$ on a single round-trip of distance $2L$: a laser field $A^+(t)=A_0e^{-iwt}$ is emitted at position 0 (see Fig. \ref{AllerRetour}) and reflected at distance L on a vibrating mirror. 
\begin{figure}[!h]
	\centering
\includegraphics[width=7cm]{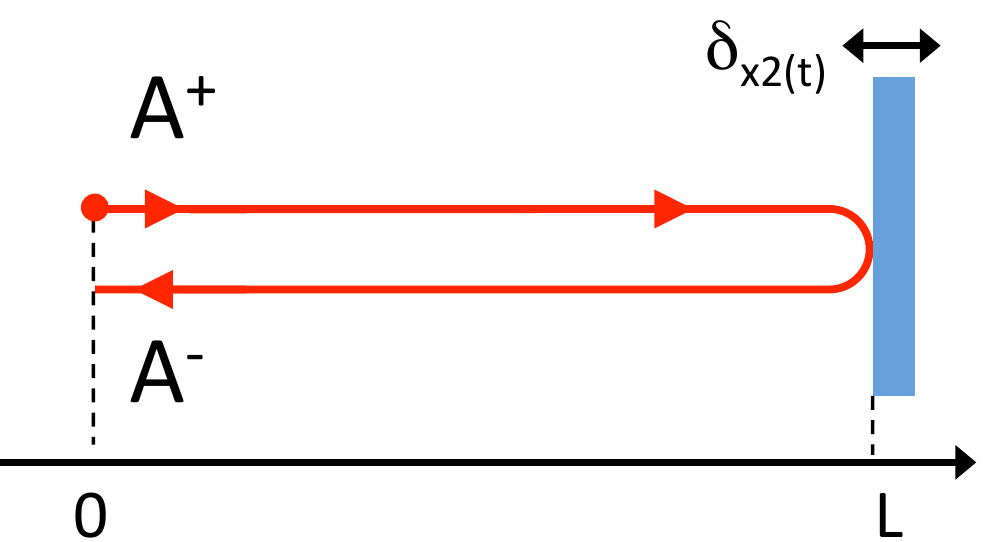}
\caption{A laser field $A^+$ is emitted at position 0 and reflected at distance L on a vibrating mirror. $A^-$ is the returning field.}
\label{AllerRetour2}
\end{figure}
When its comes back to point position 0, the field $A^-(t)$ can be expressed: 
\begin{equation}
A^-(t)= A^+(t_r)
\end{equation}
where $t_r$ is the retarded time introduced in Sec. \ref{par:cavity} (see Eq.\ref{eq:RetartedTime}) : 
\begin{equation}
t_r=t-\frac{2L}{c}-\frac{2\delta x}{c} \cos (\Omega_{x}(t-L/c)).
\label{RetardedTimeDx}
\end{equation}
The effect of a mirror vibration of amplitude $\delta x$ on a round trip is therefore completely similar to the one of a GW of amplitude $h=-\frac{2\delta x}{L}$ (see Eq.\ref{RetardedTimeDx} and \ref{RetardedTimeGW}).

If we consider a cavity formed by two mirrors placed at position $x_1$, $x_2$ (see Fig \ref{MIGAScheme2}), the effect of the vibration of a cavity mirror on the frequency noise $\delta\nu_{x}(t)$ of the resonating field can then be determined in analogy with Annex A. We obtain:
\begin{equation}
\delta\nu_{\delta x_{1,2}}(t)=\pm\frac{\Omega_{x_{1,2}}}{\omega_p }\nu_0\frac{\delta x_{1,2}}{L} \sin (\Omega_{x_{1,2}}t).
\end{equation}
The effect of both cavity mirrors on the frequency noise of the circulating field is anti-symmetric. This can be simply explained by the fact that the effect on the accumulated phase on a single round trip is opposite for the two mirrors.

\section*{Data Availability} 

The datasets studied during the current study are available from the corresponding author on reasonable request.

\section*{References}

\bibliographystyle{iopart-num}
\bibliography{MIGA}

\providecommand{\newblock}{}
\begin{thebibliography}{100}
\expandafter\ifx\csname url\endcsname\relax
  \def\url#1{{\tt #1}}\fi
\expandafter\ifx\csname urlprefix\endcsname\relax\def\urlprefix{URL }\fi
\providecommand{\eprint}[2][]{\url{#2}}

\bibitem{Carnal91}
Carnal O and Mlynek J 1991 {\em Phys.\ Rev.\ Lett.\/} {\bf 66} 2689
  \urlprefix\url{http://dx.doi.org/10.1103/PhysRevLett.66.2689}

\bibitem{Keith91}
Keith D~W, Ekstrom C~R, Turchette Q~A and Pritchard D~E 1991 {\em Phys.\ Rev.\
  Lett.\/} {\bf 66} 2693
  \urlprefix\url{http://dx.doi.org/10.1103/PhysRevLett.66.2693}

\bibitem{Riehle91}
Riehle F, Kisters T, Witte A, Helmcke J and Bord{\'e} C~J 1991 {\em Phys.\
  Rev.\ Lett.\/} {\bf 67} 177
  \urlprefix\url{http://dx.doi.org/10.1103/PhysRevLett.67.177}

\bibitem{Kasevich91}
Kasevich M and Chu S 1991 {\em Phys. Rev. Lett.\/} {\bf 67} 181
  \urlprefix\url{http://dx.doi.org/10.1103/PhysRevLett.67.181}

\bibitem{Gustavson2000}
Gustavson L, Landragin A and Kasevich M~A 2000 {\em Classical Quant. Grav.\/}
  {\bf 17} 2385 \urlprefix\url{http://dx.doi.org/10.1088/0264-9381/17/12/311}

\bibitem{Peters2001}
Peters A, Chung K~Y and Chu S 2001 {\em Metrologia\/} {\bf 38} 25
  \urlprefix\url{http://dx.doi.org/10.1088/0026-1394/38/1/4}

\bibitem{Canuel2006}
Canuel B, Leduc F, Holleville D, Gauguet A, Fils J, Virdis A, Clairon A,
  Dimarcq N, Bord{\'e} C~J, Landragin A and Bouyer P 2006 {\em Phys.\ Rev.\
  Lett.\/} {\bf 97} 010402
  \urlprefix\url{http://dx.doi.org/10.1103/PhysRevLett.97.010402}

\bibitem{Fixler2007}
Fixler J~B, Foster G~T, McGuirk J~M and Kasevich M~A 2007 {\em Science\/} {\bf
  315} 74 \urlprefix\url{http://dx.doi.org/10.1126/science.1135459}

\bibitem{Rosi2014}
Rosi G, Sorrentino F, Cacciapuoti L, Prevedelli M and Tino G 2014 {\em
  Nature\/} {\bf 510} 518 \urlprefix\url{http://dx.doi.org/10.1038/nature13433}

\bibitem{Bouchendira2011}
Bouchendira R, Clad{\'e} P, Guellati-Kh{\'e}lifa S, Nez F and Biraben F 2011
  {\em Phys.\ Rev.\ Lett.\/} {\bf 106} 080801
  \urlprefix\url{http://dx.doi.org/10.1103/PhysRevLett.106.080801}

\bibitem{Parker191}
Parker R~H, Yu C, Zhong W, Estey B and M{\"u}ller H 2018 {\em Science\/} {\bf
  360} 191--195
  \urlprefix\url{http://science.sciencemag.org/content/360/6385/191}

\bibitem{Peters99}
Peters A, Chung K~Y and Chu S 1999 {\em Nature\/} {\bf 400} 849
  \urlprefix\url{http://dx.doi.org/doi:10.1038/23655}

\bibitem{Gillot2014}
Gillot P, Francis O, Landragin A, Santos F~P~D and Merlet S 2014 {\em
  Metrologia\/} {\bf 51} L15--L17
  \urlprefix\url{http://dx.doi.org/10.1088/0026-1394/51/5/l15}

\bibitem{Freier2016}
Freier C, Hauth M, Schkolnik V, Leykauf B, Schilling M, Wziontek H, Scherneck
  H~G, M\"{u}ller J and Peters A 2016 {\em J. Phys. Conf. Ser.\/} {\bf 723}
  012050 \urlprefix\url{http://dx.doi.org/10.1088/1742-6596/723/1/012050}

\bibitem{Snadden98}
Snadden M~J, McGuirk J~M, Bouyer P, Haritos K~G and Kasevich M~A 1998 {\em
  Phys.\ Rev.\ Lett.\/} {\bf 81} 971
  \urlprefix\url{http://dx.doi.org/10.1103/PhysRevLett.81.971}

\bibitem{deAngelis2009}
de~Angelis M, Bertoldi A, Cacciapuoti L, Giorgini A, Lamporesi G, Prevedelli M,
  Saccorotti G, Sorrentino F and Tino G~M 2009 {\em Meas.\ Sci.\ Technol.\/}
  {\bf 20} 022001
  \urlprefix\url{http://dx.doi.org/10.1088/0957-0233/20/2/022001}

\bibitem{Metje2011}
Metje N, Chapman D~N, Rogers C~D~F and Bongs K 2011 {\em Adv.\ Civ.\ Eng.\/}
  {\bf 903758} \urlprefix\url{http://dx.doi.org/10.1155/2011/903758}

\bibitem{KasevichPatent}
Kasevich M~A and Dubetsky B united States Patent 7317184

\bibitem{Battelier2016}
Battelier B, Barrett B, Fouch{\'{e}} L, Chichet L, Antoni-Micollier L, Porte H,
  Napolitano F, Lautier J, Landragin A and Bouyer P 2016 Development of compact
  cold-atom sensors for inertial navigation {\em Quantum Optics\/} ed Stuhler J
  and Shields A~J ({SPIE}) \urlprefix\url{http://dx.doi.org/10.1117/12.2228351}

\bibitem{Asenbaum2017}
Asenbaum P, Overstreet C, Kovachy T, Brown D~D, Hogan J~M and Kasevich M~A 2017
  {\em Phys. Rev. Lett.\/} {\bf 118} 183602
  \urlprefix\url{https://doi.org/10.1103/physrevlett.118.183602}

\bibitem{PhysRevLett.120.043602}
Geiger R and Trupke M 2018 {\em Phys. Rev. Lett.\/} {\bf 120}(4) 043602
  \urlprefix\url{https://link.aps.org/doi/10.1103/PhysRevLett.120.043602}

\bibitem{Dimopoulos2007}
Dimopoulos S, Graham P, Hogan J and Kasevich M 2007 {\em Phys.\ Rev.\ Lett.\/}
  {\bf 98} 111102
  \urlprefix\url{http://dx.doi.org/10.1103/PhysRevLett.98.111102}

\bibitem{Geiger2011}
Geiger R, M{\'{e}}noret V, Stern G, Zahzam N, Cheinet P, Battelier B, Villing
  A, Moron F, Lours M, Bidel Y, Bresson A, Landragin A and Bouyer P 2011 {\em
  Nat. Commun.\/} {\bf 2} 474
  \urlprefix\url{http://dx.doi.org/10.1038/ncomms1479}

\bibitem{Schlippert2014}
Schlippert D, Hartwig J, Albers H, Richardson L~L, Schubert C, Roura A,
  Schleich W~P, Ertmer W and Rasel E~M 2014 {\em Phys.\ Rev.\ Lett.\/} {\bf
  112} 203002 \urlprefix\url{http://dx.doi.org/10.1103/PhysRevLett.112.203002}

\bibitem{Barrett2015}
Barrett B, Antoni-Micollier L, Chichet L, Battelier B, Gominet P, Bertoldi A,
  Bouyer P and Landragin A 2015 {\em New J. Phys.\/} {\bf 17} 085010
  \urlprefix\url{http://dx.doi.org/10.1088/1367-2630/17/8/085010}

\bibitem{Barrett2016}
Barrett B, Bertoldi A and Bouyer P Inertial quantum sensors using light and
  matter arXiv:1603.03246 [physics.atom-ph]

\bibitem{Rosi2017}
Rosi G, D'Amico G, Cacciapuoti L, Sorrentino F, Prevedelli M, Zych M, Brukner
  {\v{C}} and Tino G~M 2017 {\em Nat. Commun.\/} {\bf 8} 15529
  \urlprefix\url{http://dx.doi.org/10.1038/ncomms15529}

\bibitem{Muller2008}
M{\"u}ller H, Chiow S~w, Herrmann S, Chu S and Chung K~Y 2008 {\em Phys.\ Rev.\
  Lett.\/} {\bf 100} 031101
  \urlprefix\url{http://dx.doi.org/10.1103/PhysRevLett.100.031101}

\bibitem{Jentsch2004}
Jentsch C, M{\"u}ller T, Rasel E~M and Ertmer W 2004 {\em Gen. Rel. Gravit.\/}
  {\bf 36} 2197
  \urlprefix\url{http://dx.doi.org/10.1023/B%3AGERG.0000046179.26175.fa}

\bibitem{Aguilera2014}
Aguilera D {\em et~al.\/} 2014 {\em Classical Quant. Grav.\/} {\bf 31} 115010
  \urlprefix\url{http://dx.doi.org/10.1088/0264-9381/31/11/115010}

\bibitem{Schubert2013}
Schubert C {\em et~al.\/} Differential atom interferometry with $^{87}${R}b and
  $^{85}${R}b for testing the {WEP} in {STE}-{QUEST} (\textit{Preprint}
  \eprint{arXiv:1312.5963 [physics.atom-ph]})

\bibitem{Tino2013}
Tino G~M {\em et~al.\/} 2013 {\em Nucl.\ Phys.\ B - Proc. Suppl.\/} {\bf
  243-244} 203
  \urlprefix\url{http://dx.doi.org/10.1016/j.nuclphysbps.2013.09.023}

\bibitem{Borde1983}
Borde C~J, Sharma J, Tourrenc P and Damour T 1983 {\em J. Physique\/} {\bf 44}
  983--990
  \urlprefix\url{http://dx.doi.org/10.1051/jphyslet:019830044024098300}

\bibitem{Chiao2004}
Chiao R~Y and Speliotopoulos A~D 2004 {\em J.\ Mod.\ Opt.\/} {\bf 51} 861
  \urlprefix\url{http://dx.doi.org/10.1080/09500340408233603}

\bibitem{Foffa2006}
Foffa S, Gasparini A, Papucci M and Sturani R 2006 {\em Phys.\ Rev.\ D\/} {\bf
  73} 022001 \urlprefix\url{http://dx.doi.org/10.1103/PhysRevD.73.022001}

\bibitem{Tino2007}
Tino G~M and Vetrano F 2007 {\em Classical Quant. Grav.\/} {\bf 24} 2167
  \urlprefix\url{http://dx.doi.org/10.1088/0264-9381/24/9/001}

\bibitem{GWdet}
{LIGO Scientific Collaboration and Virgo Collaboration} 2016 {\em Phys. Rev.
  Lett.\/} {\bf 116} 061102
  \urlprefix\url{https://doi.org/10.1103/PhysRevLett.116.061102}

\bibitem{Sesana}
Sesana A 2016 {\em Phys. Rev. Lett.\/} {\bf 116} 231102
  \urlprefix\url{https://doi.org/10.1103/PhysRevLett.116.231102}

\bibitem{Graham2018}
Graham P~W and Jung S 2018 {\em Phys. Rev. D\/} {\bf 97}
  \urlprefix\url{http://dx.doi.org/10.1103/physrevd.97.024052}

\bibitem{Abbott2017}
Abbott B~P {\em et~al.\/} 2017 {\em Astrophys. J\/} {\bf 848} L12
  \urlprefix\url{https://doi.org/10.3847/2041-8213/aa91c9}

\bibitem{Christensen}
{NL Christensen for the LIGO Scientific Collaboration and the Virgo
  Collaboration} Multimessenger astronomy (\textit{Preprint}
  \eprint{arXiv:1105.5843})

\bibitem{Mandel2018}
Mandel I, Sesana A and Vecchio A 2018 {\em Classical Quant. Grav.\/} {\bf 35}
  054004 \urlprefix\url{https://doi.org/10.1088/1361-6382/aaa7e0}

\bibitem{Dimopoulos2009}
Dimopoulos S, Graham P~W, Hogan J~M, Kasevich M~A and Rajendran S 2009 {\em
  Phys.\ Lett.\ B\/} {\bf 678} 37
  \urlprefix\url{http://dx.doi.org/10.1016/j.physletb.2009.06.011}

\bibitem{Dimopoulos2008}
Dimopoulos S, Graham P, Hogan J, Kasevich M and Rajendran S 2008 {\em Phys.\
  Rev.\ D\/} {\bf 78} 122002
  \urlprefix\url{http://dx.doi.org/10.1103/PhysRevD.78.122002}

\bibitem{Harms2013}
Harms J, Slagmolen B, Adhikari R, Miller M, Evans M, Chen Y, M{\"u}ller H and
  Ando M 2013 {\em Phys.\ Rev.\ D\/} {\bf 88} 122003
  \urlprefix\url{http://dx.doi.org/10.1103/PhysRevD.88.122003}

\bibitem{Canuel2014}
Canuel B, Amand L, Bertoldi A, Chaibi W, Geiger R, Gillot J, Landragin A,
  Merzougui M, Riou I, Schmid S~P and Bouyer P 2014 {\em E3S Web of
  Conferences\/} {\bf 4} 01004
  \urlprefix\url{http://dx.doi.org/10.1051/e3sconf/20140401004}

\bibitem{Canuel2016}
Canuel B, Pelisson S, Amand L, Bertoldi A, Cormier E, Fang B, Gaffet S, Geiger
  R, Harms J, Holleville D, Landragin A, Lef{\'e}vre G, Lhermite J, Mielec N,
  Prevedelli M, Riou I and Bouyer P 2016 {\em Proc. SPIE\/} {\bf 9900} 990008
  \urlprefix\url{http://dx.doi.org/10.1117/12.22288251}

\bibitem{Hamilton2015}
Hamilton P, Jaffe M, Brown J~M, Maisenbacher L, Estey B and M{\"u}ller H 2015
  {\em Phys.\ Rev.\ Lett.\/} {\bf 114} 100405
  \urlprefix\url{http://dx.doi.org/10.1103/PhysRevLett.114.100405}

\bibitem{Riou2017}
Riou I, Mielec N, Lef{\`{e}}vre G, Prevedelli M, Landragin A, Bouyer P,
  Bertoldi A, Geiger R and Canuel B 2017 {\em J. Phys. B\/} {\bf 50} 155002
  \urlprefix\url{http://dx.doi.org/10.1088/1361-6455/aa7592}

\bibitem{Cheinet08}
Cheinet P, Canuel B, {Pereira Dos Santos} F, Gauguet A, Yver-Leduc F and
  Landragin A 2008 {\em IEEE T.\ Instrum.\ Meas.\/} {\bf 57} 1141
  \urlprefix\url{http://dx.doi.org/10.1109/TIM.2007.915148}

\bibitem{Zhu1993}
Zhu M and Hall J~L 1993 {\em J. Opt. Soc. Am. B\/} {\bf 10} 802
  \urlprefix\url{https://doi.org/10.1364/josab.10.000802}

\bibitem{Numata2004}
Numata K, Kemery A and Camp J 2004 {\em Phys. Rev. Lett.\/} {\bf 93}(25) 250602
  \urlprefix\url{http://link.aps.org/doi/10.1103/PhysRevLett.93.250602}

\bibitem{Thorpe2011}
Thorpe M~J, Rippe L, Fortier T~M, Kirchner M~S and Rosenband T 2011 {\em Nature
  Photon.\/} {\bf 5} 688--693
  \urlprefix\url{http://dx.doi.org/10.1038/nphoton.2011.215}

\bibitem{Jiang2011}
Jiang Y~Y, Ludlow A~D, Lemke N~D, Fox R~W, Sherman J~A, Ma L~S and Oates C~W
  2011 {\em Nat. Photon.\/} {\bf 5} 158--161
  \urlprefix\url{http://dx.doi.org/10.1038/nphoton.2010.313}

\bibitem{Saulson}
Saulson P 1984 {\em Phys. Rev. D\/} {\bf 30} 732--736
  \urlprefix\url{https://doi.org/10.1103/PhysRevD.30.732}

\bibitem{Harms2015}
Harms J 2015 {\em Living Rev. Relativity\/} {\bf 18} 3
  \urlprefix\url{https://doi.org/10.1007/lrr-2015-3}

\bibitem{Chaibi2016}
Chaibi W, Geiger R, Canuel B, Bertoldi A, Landragin A and Bouyer P 2016 {\em
  Phys. Rev. D\/} {\bf 93} 021101
  \urlprefix\url{http://dx.doi.org/10.1103/PhysRevD.93.021101}

\bibitem{Abend2016}
Abend S, Gebbe M, Gersemann M, Ahlers H, M\"{u}ntinga H, Giese E, Gaaloul N,
  Schubert C, L\"{a}mmerzahl C, Ertmer W, Schleich W and Rasel E 2016 {\em
  Phys. Rev. Lett.\/} {\bf 117}
  \urlprefix\url{https://doi.org/10.1103/physrevlett.117.203003}

\bibitem{McDonald2013a}
McDonald G, Keal H, Altin P, Debs J, Bennetts S, Kuhn C, Hardman K, Johnsson M,
  Close J and Robins N 2013 {\em Phys.\ Rev.\ A\/} {\bf 87} 013632
  \urlprefix\url{http://dx.doi.org/10.1103/PhysRevA.87.013632}

\bibitem{McDonald2013}
McDonald G~D, Kuhn C~C, Bennetts S, Debs J~E, Hardman K~S, Johnsson M, Close
  J~D and Robins N~P 2013 {\em Phys.\ Rev.\ A\/} {\bf 88} 053620
  \urlprefix\url{http://dx.doi.org/10.1103/PhysRevA.88.053620}

\bibitem{Hosten2016}
Hosten O, Engelsen N~J, Krishnakumar R and Kasevich M~A 2016 {\em Nature\/}
  {\bf 529} 505--508 \urlprefix\url{https://doi.org/10.1038/nature16176}

\bibitem{Cox2016}
Cox K~C, Greve G~P, Weiner J~M and Thompson J~K 2016 {\em Phys. Rev. Lett.\/}
  {\bf 116} \urlprefix\url{https://doi.org/10.1103/physrevlett.116.093602}

\bibitem{Graham2013}
Graham P~W, Hogan J~M, Kasevich M~A and Rajendran S 2013 {\em Phys.\ Rev.\
  Lett.\/} {\bf 110} 171102
  \urlprefix\url{http://dx.doi.org/10.1103/PhysRevLett.110.171102}

\bibitem{Santarelli1999}
Santarelli G, Laurent P, Lemonde P, Clairon A, Mann A~G, Chang S, Luiten A~N
  and Salomon C 1999 {\em Phys. Rev. Lett.\/} {\bf 82} 4619--4622
  \urlprefix\url{http://dx.doi.org/10.1103/physrevlett.82.4619}

\bibitem{Rocco2014}
Rocco E, Palmer R~N, Valenzuela T, Boyer V, Freise A and Bongs K 2014 {\em New
  J.\ Phys.\/} {\bf 16} 093046
  \urlprefix\url{http://dx.doi.org/10.1088/1367-2630/16/9/093046}

\bibitem{MUQUANS}
{h}ttp://www.muquans.com/

\bibitem{Chiow2012}
Chiow S~w, Kovachy T, Hogan J~M and Kasevich M~A 2012 {\em Opt.\ Lett.\/} {\bf
  37} 3861 \urlprefix\url{http://dx.doi.org/10.1364/OL.37.003861}

\bibitem{San2012}
San{\'{e}} S~S, Bennetts S, Debs J~E, Kuhn C~C~N, McDonald G~D, Altin P~A,
  Close J~D and Robins N~P 2012 {\em Opt. Expr.\/} {\bf 20} 8915
  \urlprefix\url{https://doi.org/10.1364/oe.20.008915}

\bibitem{Fang2018}
Fang B, Mielec N, Savoie D, Altorio M, Landragin A and Geiger R 2018 {\em New
  J. Phys.\/} {\bf 20} 023020
  \urlprefix\url{http://dx.doi.org/10.1088/1367-2630/aaa37c}

\bibitem{DovaleAlvarez:2017}
Dovale-{\'A}lvarez M, Brown D~D, Jones A~W, Mow-Lowry C~M, Miao H and Freise A
  2017 {\em Physical Review A\/} {\bf 96} 053820--10
  \urlprefix\url{http://dx.doi.org/10.1103/PhysRevA.96.053820}

\bibitem{Kovachy2015b}
Kovachy T, Asenbaum P, Overstreet C, Donnelly C~A, Dickerson S~M, Sugarbaker A,
  Hogan J~M and Kasevich M~A 2015 {\em Nature\/} {\bf 528} 530--533
  \urlprefix\url{doi:10.1038/nature16155}

\bibitem{Bertoldi2006}
Bertoldi A, Lamporesi G, Cacciapuoti L, De~Angelis M, Fattori M, Petelski T,
  Peters A, Prevedelli M, Stuhler J and Tino G 2006 {\em Eur.\ Phys.\ J.\ D\/}
  {\bf 40} 271--279
  \urlprefix\url{http://dx.doi.org/10.1140/epjd/e2006-00212-2}

\bibitem{Steck2001}
Steck D~A 2001 Rubidium 87 d line data
  \urlprefix\url{http://steck.us/alkalidata/rubidium87numbers.1.6.pdf}

\bibitem{Dickerson2012}
Dickerson S, Hogan J~M, Johnson D~M~S, Kovachy T, Sugarbaker A, wey Chiow S and
  Kasevich M~A 2012 {\em Rev. Sci. Instrum.\/} {\bf 83} 065108
  \urlprefix\url{https://doi.org/10.1063/1.4720943}

\bibitem{Schuldt2015}
Schuldt T {\em et~al.\/} 2015 {\em Exp.\ Astronom.\/}  1--40 ISSN 0922-6435
  \urlprefix\url{http://dx.doi.org/10.1007/s10686-014-9433-y}

\bibitem{KubelkaLange2016}
Kubelka-Lange A, Herrmann S, Grosse J, L\"{a}mmerzahl C, Rasel E~M and
  Braxmaier C 2016 {\em Rev. Sci. Instrum.\/} {\bf 87} 063101
  \urlprefix\url{https://doi.org/10.1063/1.4952586}

\bibitem{Bettini2012}
Bettini A 2012 {\em Eur.\ Phys.\ J.-Plus.\/} {\bf 127} 114
  \urlprefix\url{http://dx.doi.org/10.1140/epjp/i2012-12114-y}

\bibitem{Gaffet2009}
Gaffet S, Wang J, Yedlin M, Nolet G, Maron C, Brunel D, Cavaillou A, Boyer D,
  Sudre C and Auguste M 2009 {\em IRIS Data Services Newsletter\/} {\bf 11} 3
  \urlprefix\url{http://ds.iris.edu/ds/newsletter/vol11/no3/a-3d-broadband-seismic-array-at-lsbb/}

\bibitem{Ford2007}
Ford D and Williams P 2007 {\em Karst Hydrogeology and Geomorphology\/} (John
  Wiley {\&} Sons Ltd) \urlprefix\url{https://doi.org/10.1002/9781118684986}

\bibitem{Garry2008}
Garry B, Blondel T, Emblanch C, Sudre C, Bilgot S, Cavaillou A, Boyer D and
  Auguste M 2008 {\em Int. J. Spel.\/} {\bf 37} 75
  \urlprefix\url{http://dx.doi.org/10.5038/1827-806X.37.1.7}

\bibitem{networkHydro}
Network of hydrogeological research sites http://hplus.ore.fr/

\bibitem{Carrire2016}
Carri{\`{e}}re S~D, Chalikakis K, Danquigny C, Davi H, Mazzilli N, Ollivier C
  and Emblanch C 2016 {\em Hydrogeol. J.\/} {\bf 24} 1905
  \urlprefix\url{https://doi.org/10.1007/s10040-016-1425-8}

\bibitem{Lesparre2016}
Lesparre N, Boudin F, Champollion C, Ch{\'{e}}ry J, Danquigny C, Seat H~C,
  Cattoen M, Lizion F and Longuevergne L 2016 {\em Geophys. J. Int.\/} {\bf
  208} 1389 \urlprefix\url{https://doi.org/10.1093/gji/ggw446}

\bibitem{Wang2010}
Wang J, Guglielmi Y and Gaffet S 2010 Collaborative projects between two
  {USA-F}rance national subsurface laboratories to improve imaging of
  fractured-porous rocks properties and coupled {THMCB} processes {\em Rock
  Mechanics in Civil and Environmental Engineering\/} ed Dudt and Mathier
  (London: Taylor and Francis Group) pp 857--860 ISBN 978-0-415-58654-2

\bibitem{Deville2012}
Deville S, Jacob T, Chery J and Champollion C 2012 {\em Geophys. J. Int.\/}
  {\bf 192} 82 \urlprefix\url{https://doi.org/10.1093/gji/ggs007}

\bibitem{Fores2016}
Fores B, Champollion C, Moigne N~L, Bayer R and Ch{\'{e}}ry J 2016 {\em
  Geophys. J. Int.\/} {\bf 208} 269--280
  \urlprefix\url{https://doi.org/10.1093/gji/ggw396}

\bibitem{Gaffet2003}
Gaffet S, Guglielmi Y, Virieux J, Waysand G, Chwala A, Stolz R, Emblanch C,
  Auguste M, Boyer D and Cavaillou A 2003 {\em Geophys.\ J.\ Int.\/} {\bf 155}
  981 \urlprefix\url{http://dx.doi.org/10.1111/j.1365-246X.2003.02095.x}

\bibitem{Zandi2011}
Zandi A~S, Dumont G~A, Yedlin M~J, Lapeyrie P, Sudre C and Gaffet S 2011 {\em
  {IEEE} Trans. Biomed. Eng\/} {\bf 58} 2407--2417
  \urlprefix\url{https://doi.org/10.1109/tbme.2011.2158647}

\bibitem{Farah2014}
Farah T, Guerlin C, Landragin A, Bouyer P, Gaffet S, Santos F~P~D and Merlet S
  2014 {\em Gyroscopy and Navigation\/} {\bf 5} 266
  \urlprefix\url{http://dx.doi.org/10.1134/S2075108714040051}

\bibitem{Peterson1993}
Peterson J 1993 Observations and modelling of seismic background noise {U}.{S}.
  Geol. Surv. Open-File Rept. 93-332, Albuquerque, New Mexico (1993)
  \urlprefix\url{http://earthquake.usgs.gov/regional/asl/pubs/files/ofr93-322.pdf}

\bibitem{Rosat2016}
Rosat S, Hinderer J, Boy J~P, Littel F, Boyer D, Bernard J~D, Rogister Y,
  M{\'e}min A and Gaffet S 2016 {\em E3S Web of Conf.\/} {\bf 12} 06003
  \urlprefix\url{https://doi.org/10.1051/e3sconf/20161206003}

\bibitem{Rosat2011}
Rosat S and Hinderer J 2011 {\em Bull.\ Seism.\ Soc.\ Am.\/} {\bf 101} 1233
  \urlprefix\url{http://dx.doi.org/10.1785/0120100217}

\bibitem{Berger2004}
Berger J, Davis P and Ekström G 2004 {\em J. Geophys. Res.\/} {\bf 109} B11307
  \urlprefix\url{http://dx.doi.org/10.1029/2004JB003408}

\bibitem{Canuel2007}
Canuel B 2007 Etude d'un gyrom{\`e}tre {\`a} atomes froids ph{D} Thesis,
  Universit{\'e} Paris XI
  \urlprefix\url{https://tel.archives-ouvertes.fr/tel-00193288}

\bibitem{Waysand2000}
Waysand G, Bloyet D, Bongiraud J, Collar J, Dolabdjian C and Thiec P~L 2000
  {\em Instrum.\ Meth.\ A\/} {\bf 444} 336
  \urlprefix\url{http://dx.doi.org/doi:10.1016/S0168-9002(99)01377-7}

\bibitem{Henry2016}
Henry S, {Pozzo di Borgo} E, Danquigny C and Abi B 2016 {\em E3S Web of
  Conferences\/} {\bf 12} 02003
  \urlprefix\url{https://doi.org/10.1051/e3sconf/20161202003}

\bibitem{Henry2014}
Henry S, {Pozzo di Borgo} E, Danquigny C, Cavaillou A, Cottle A, Gaffet S and
  Pipe M 2014 {\em E3S Web of Conferences\/} {\bf 4} 02004
  \urlprefix\url{http://dx.doi.org/10.1051/e3sconf/20140402004}

\bibitem{Sorrentino2012}
Sorrentino F, Bertoldi A, Bodart Q, Cacciapuoti L, de~Angelis M, Lien Y~H,
  Prevedelli M, Rosi G and Tino G~M 2012 {\em Appl.\ Phys.\ Lett.\/} {\bf 101}
  114106 \urlprefix\url{http://dx.doi.org/10.1063/1.4751112}

\bibitem{Rosi2015}
Rosi G, Cacciapuoti L, Sorrentino F, Menchetti M, Prevedelli M and Tino G~M
  2015 {\em Phys.\ Rev.\ Lett.\/} {\bf 114} 013001
  \urlprefix\url{http://dx.doi.org/10.1103/PhysRevLett.114.013001}

\bibitem{Lautier2014}
Lautier J, Volodimer L, Hardin T, Merlet S, Lours M, Santos F~P~D and Landragin
  A 2014 {\em Appl.\ Phys.\ Lett.\/} {\bf A05} 144102
  \urlprefix\url{http://dx.doi.org/10.1063/1.4897358}

\bibitem{Gray1995}
Gray S~D, Parmentola J~A and LeSchack R 1995 {\em J.\ Phys.\ D Appl.\ Phys.\/}
  {\bf 28} 2378 \urlprefix\url{http://dx.doi.org/10.1088/0022-3727/28/11/024}

\bibitem{Lamporesi2007}
Lamporesi G, Bertoldi A, Cecchetti A, Duhlach B, Fattori M, Malengo A,
  Pettorruso S, Prevedelli M and Tino G 2007 {\em Rev.\ Sci.\ Instrum.\/} {\bf
  78} 075109 \urlprefix\url{http://dx.doi.org/10.1063/1.2751090}

\bibitem{VolMoody2002}
Moody M~V, Paik H~J and Canavan E~R 2002 {\em Rev.\ Sci.\ Instrum.\/} {\bf 73}
  3957--3974 \urlprefix\url{http://dx.doi.org/10.1063/1.1511798}

\bibitem{Butler1984}
Butler D~K 1984 {\em Geophys.\/} {\bf 49} 1084
  \urlprefix\url{http://dx.doi.org/10.1190/1.1441723}

\bibitem{Berkowitz2002}
Berkowitz B 2002 {\em Adv. Water Resour.\/} {\bf 25} 861
  \urlprefix\url{https://doi.org/10.1016/s0309-1708(02)00042-8}

\bibitem{Chalikakis2011}
Chalikakis K, Plagnes V, Guerin R, Valois R and Bosch F~P 2011 {\em Hydrogeol.
  J.\/} {\bf 19} 1169--1180
  \urlprefix\url{https://doi.org/10.1007/s10040-011-0746-x}

\bibitem{Carrire2013}
Carri{\`{e}}re S~D, Chalikakis K, S{\'{e}}n{\'{e}}chal G, Danquigny C and
  Emblanch C 2013 {\em J. Appl. Geophys.\/} {\bf 94} 31
  \urlprefix\url{https://doi.org/10.1016/j.jappgeo.2013.03.014}

\bibitem{Hector2014}
Hector B, Hinderer J, S{\'{e}}guis L, Boy J~P, Calvo M, Descloitres M, Rosat S,
  Galle S and Riccardi U 2014 {\em J. Geodyn.\/} {\bf 80} 34
  \urlprefix\url{https://doi.org/10.1016/j.jog.2014.04.003}

\bibitem{Hasan2008}
Hasan S, Troch P~A, Bogaart P~W and Kroner C 2008 {\em Water Resour. Res\/}
  {\bf 44} \urlprefix\url{https://doi.org/10.1029/2007wr006321}

\bibitem{Pool1995}
Pool D~R and Eychaner J~H 1995 {\em Ground Water\/} {\bf 33} 425
  \urlprefix\url{https://doi.org/10.1111/j.1745-6584.1995.tb00299.x}

\bibitem{Geiger2015}
Geiger R, Amand L, Bertoldi A, Canuel B, Chaibi W, Danquigny C, Dutta I, Fang
  B, Gaffet S, Gillot J, Holleville D, Landragin A, Merzougui M, Riou I, Savoie
  D and Bouyer P 2015 Matter-wave laser interferometric gravitation antenna
  ({MIGA}): {N}ew perspectives for fundamental physics and geosciences {\em
  Proceedings of the 50th Rencontres de Moriond {"}100 years after GR{"}, La
  Thuile (Italy), 21-28 March 2015\/} ed Augé E, Dumarchez J and Trân
  Thanh~Vân J (ARSIF) \urlprefix\url{http://arxiv.org/abs/1505.07137}

\bibitem{Ghasemizadeh2012}
Ghasemizadeh R, Hellweger F, Butscher C, Padilla I, Vesper D, Field M and
  Alshawabkeh A 2012 {\em Hydrogeol. J.\/} {\bf 20} 1441
  \urlprefix\url{https://doi.org/10.1007/s10040-012-0897-4}

\bibitem{Thompson2007}
Thompson A~R, Moran J~M and Swenson G~W 2007 {\em Interferometry and Synthesis
  in Radio Astronomy, Second Edition\/} (John Wiley \& Sons, inc.) ISBN
  9780471254928 \urlprefix\url{http://dx.doi.org/doi:10.1002/9783527617845}

\bibitem{Greenaway1991}
Greenaway A~H 1991 {\em Meas.\ Sci.\ Technol.\/} {\bf 2} 1
  \urlprefix\url{http://dx.doi.org/doi:10.1088/0957-0233/2/1/001}

\bibitem{Chiow2011}
Chiow S~w, Kovachy T, Chien H~C and Kasevich M~A 2011 {\em Phys.\ Rev.\
  Lett.\/} {\bf 107} 130403
  \urlprefix\url{http://dx.doi.org/10.1103/PhysRevLett.107.130403}

\bibitem{Estey2015}
Estey B, Yu C, M{\"u}ller H, Kuan P~C and Lan S~Y 2015 {\em Phys.\ Rev.\
  Lett.\/} {\bf 115} 083002
  \urlprefix\url{http://dx.doi.org/10.1103/PhysRevLett.115.083002}

\bibitem{Dutta2016}
Dutta I, Savoie D, Fang B, Venon B, Alzar C~G, Geiger R and Landragin A 2016
  {\em Phys. Rev. Lett.\/} {\bf 116}
  \urlprefix\url{http://dx.doi.org/10.1103/physrevlett.116.183003}

\bibitem{Kohlhaas2015}
Kohlhaas R, Bertoldi A, Cantin E, Aspect A, Landragin A and Bouyer P 2015 {\em
  Phys.\ Rev.\ X\/} {\bf 5} 021011
  \urlprefix\url{http://dx.doi.org/10.1103/PhysRevX.5.021011}

\bibitem{VPB2006}
{The VIRGO Collaboration} 2006 The virgo physics book, vol. ii
  \urlprefix\url{https://www.ego-gw.it/public/events/vesf/2010/Presentations/Interferometer_Materials-Vinet.pdf}

\end{thebibliography}

\section*{Acknowledgments} 

This work was realized with the financial support of the French State through the ``Agence Nationale de la Recherche'' (ANR) in the frame of the ``Investissement d'avenir'' programs: Equipex MIGA (ANR-11-EQPX-0028), IdEx Bordeaux - LAPHIA (ANR-10-IDEX-03-02) and FIRST-TF (ANR-10-LABX-48-01). This work was also supported by the région d’Aquitaine (project IASIG-3D) and by the city of Paris (Emergence project HSENS-MWGRAV). We also aknowledge support from the CPER LSBB2020 project; funded by the ``région PACA'', the ``département du Vaucluse'', the MIGA Equipex and the ``FEDER PA0000321 programmation 2014-2020''. We also thank the ``Pôle de compétitivité Route des lasers-- Bordeaux'' cluster for his support. G.L. thanks DGA for financial support. M.P. thanks LAPHIA--IdEx Bordeaux for partial financial support. M.D.A. and A.F. acknowledge financial support of the Defence Science and Technology Laboratory (DSTL) and the UK National Quantum Technology Hub in Sensors and Metrology with EPSRC Grant No. EP/M013294/1. A.F. was supported by the Science and Technology Facilities Council Consolidated Grant (No. ST/N000633/1).

\section*{Author contributions}

B.C. and A.B. wrote the manuscript. B.C., A.B., R.G., M.P., S.R., W.C., S.P. and P.B. contributed to develop the model of the instrument response and the noise projections. L.A., T.C., J.G., D.H., M.M. contributed to the design of the atom source, vacuum system and mirror suspensions. B.C., L.A., B.F., R.G., J.G., D.H., J.J., G.L., N.M., I.R. and A.L. participated to assembly and preliminary experiments with the atom source. C.D., S.G. and T.M. contributed to geophysics applications of MIGA. M.D.A. and A.F. participated to interrogation cavity design. E.C. contributed to interrogation laser design. S.G. contributed to seismic noise analysis and design of infrastructure works. J.H, S.R., Y.R. contributed to the analysis of site gravimetry using the SG instrument. E.P.B. and S.H. participated to the analysis of the electromagnetic noise of the site. P.B. supervised the work and the design of the antenna. All authors reviewed the manuscript. 

\section*{Additional information} 

\textbf{Competing interests}: The authors declare no competing interests.

\end{document}